\documentclass[aps,prb ,reprint,superscriptaddress]{revtex4-1}

\usepackage{graphicx}
\usepackage{epstopdf}
\usepackage{dcolumn}
\usepackage{bm}
\usepackage{amsmath}
\usepackage{hyperref}
\hypersetup{backref, colorlinks=true, linkcolor=blue, citecolor=blue, urlcolor=blue}
\usepackage{sistyle} 
\usepackage{multirow}
\usepackage{xcolor}

\draft

\begin{document}


\title {Single shot time-resolved magnetic x-ray absorption at a Free Electron Laser}


\author{Emmanuelle Jal}
\affiliation{Sorbonne Universit\'e, CNRS, Laboratoire de Chimie Physique - Mati\`ere et Rayonnement, LCPMR, 75005 Paris, France}

\author{Mikako Makita}
\affiliation{Paul Scherrer Institut, 5232 Villigen PSI, Switzerland}

\author{Benedikt R\"osner}
\affiliation{Paul Scherrer Institut, 5232 Villigen PSI, Switzerland}

\author{Christian David}
\affiliation{Paul Scherrer Institut, 5232 Villigen PSI, Switzerland}

\author{Frithjof Nolting}
\affiliation{Paul Scherrer Institut, 5232 Villigen PSI, Switzerland}

\author{J\"org Raabe}
\affiliation{Paul Scherrer Institut, 5232 Villigen PSI, Switzerland}

\author{Tatiana Savchenko}
\affiliation{Paul Scherrer Institut, 5232 Villigen PSI, Switzerland}

\author{Armin Kleibert}
\affiliation{Paul Scherrer Institut, 5232 Villigen PSI, Switzerland}

\author{Flavio Capotondi}
\affiliation{FERMI, Elettra-Sincrotrone Trieste, SS 14 - km 163.5, 34149 Basovizza, Trieste, Italy}

\author{Emanuele Pedersoli}
\affiliation{FERMI, Elettra-Sincrotrone Trieste, SS 14 - km 163.5, 34149 Basovizza, Trieste, Italy}

\author{Lorenzo Raimondi}
\affiliation{FERMI, Elettra-Sincrotrone Trieste, SS 14 - km 163.5, 34149 Basovizza, Trieste, Italy}

\author{Michele Manfredda}
\affiliation{FERMI, Elettra-Sincrotrone Trieste, SS 14 - km 163.5, 34149 Basovizza, Trieste, Italy}

\author{Ivaylo Nikolov}
\affiliation{FERMI, Elettra-Sincrotrone Trieste, SS 14 - km 163.5, 34149 Basovizza, Trieste, Italy}

\author{Xuan Liu }
\affiliation{Sorbonne Universit\'e, CNRS, Laboratoire de Chimie Physique - Mati\`ere et Rayonnement, LCPMR, 75005 Paris, France}

\author{Alaa el dine Merhe}
\affiliation{Sorbonne Universit\'e, CNRS, Laboratoire de Chimie Physique - Mati\`ere et Rayonnement, LCPMR, 75005 Paris, France}

\author{Nicolas Jaouen}
\affiliation{Synchrotron SOLEIL, L\textquoteright Orme des Merisiers, Saint-Aubin, 91192 Gif-sur-Yvette, France}

\author{Jon Gorchon}
\affiliation{Institut Jean Lamour, Universit\'e Henri Poincar\'e, Nancy, France}

\author{Gregory Malinowski}
\affiliation{Institut Jean Lamour, Universit\'e Henri Poincar\'e, Nancy, France}

\author{Michel Hehn}
\affiliation{Institut Jean Lamour, Universit\'e Henri Poincar\'e, Nancy, France}

\author{Boris Vodungbo}
\affiliation{Sorbonne Universit\'e, CNRS, Laboratoire de Chimie Physique - Mati\`ere et Rayonnement, LCPMR, 75005 Paris, France}

\author{Jan L\"uning}
\affiliation{Sorbonne Universit\'e, CNRS, Laboratoire de Chimie Physique - Mati\`ere et Rayonnement, LCPMR, 75005 Paris, France}
\affiliation{Synchrotron SOLEIL, L\textquoteright Orme des Merisiers, Saint-Aubin, 91192 Gif-sur-Yvette, France}


\date{\today}

\begin{abstract}
Ultrafast dynamics are generally investigated using stroboscopic pump-probe measurements, which characterize the sample properties for a single, specific time delay. These measurements are then repeated for a series of discrete time delays to reconstruct the overall time trace of the process. As a consequence, this approach is limited to the investigation of fully reversible phenomena. We recently introduced an off-axis zone plate based X-ray streaking technique, which overcomes this limitation by sampling the relaxation dynamics with a single femtosecond X-ray pulse streaked over a picosecond long time window. In this article we show that the X-ray absorption cross section can be employed as the contrast mechanism in this novel technique. We show that changes of the absorption cross section on the percent level can be resolved with this method. To this end we measure the ultrafast magnetization dynamics in CoDy alloy films. Investigating different chemical compositions and infrared pump fluences, we demonstrate the routine applicability of this technique. Probing in transmission the average magnetization dynamics of the entire film, our experimental findings indicate that the demagnetization time is independent of the specific infrared laser pump fluence. These results pave the way for the investigation of irreversible phenomena in a wide variety of scientific areas.
\end{abstract}


\maketitle


\subsection{Introduction}
Femtosecond time resolved experiments employing advanced X-ray probe techniques have been realized in a wide variety of scientific domains since the advent of ultra-short-X-ray-pulse sources such as femtoslicing, high harmonic generation and X-ray Free Electron Lasers (XFELs). Research activities range from the investigation of fundamental processes in gas phase experiments \cite{squibb_acetylacetone_2018, rudenko_femtosecond_2017, young_femtosecond_2010} to the observation of structural rearrangement in biological macromolecules using novel time resolved crystallography techniques \cite{coquelle_chromophore_2017, tayeb-fligelman_cytotoxic_2017}. In condensed matter physics, the availability of ultrashort X-ray pulses has allowed to probe ultrafast charge, spin and lattice dynamics with chemical selectivity and nanometer spatial resolution \cite{ graves_nanoscale_2013, radu_transient_2011, vodungbo_indirect_2016, gerber_femtosecond_2017, chaix_dispersive_2017}. 

A significant number of these time resolved experiments investigate the relaxation dynamics following an externally induced excitation, which is typically realized by an ultrashort infrared (IR) pump pulse. To characterize these dynamics, the probe signal is generally obtained by accumulating statistics over repetitive pump-probe cycles for a specific delay. The advent of XFELs promised to overcome this need of signal accumulation due to their high pulse intensity, their temporal coherence, their femtosecond pulse duration and due to the strong sensitivity of X-ray probe techniques. The feasibility of such X-ray single-shot probing has indeed been demonstrated\cite{wang_femtosecond_2012}. Nevertheless, the pump-probe cycle has to be repeated for different time delays to reconstruct the overall time evolution of the relaxation dynamics from these individual measurements. As a consequence, identical experimental conditions must be reestablished for each pump-probe cycle, which may be compromised due to practical aspects such as the reproducibility of the pump and probe parameters. In addition, temporal jitter between pump and probe pulses limits the achievable time resolution in such experiments. Finally, a fundamental restriction concerns the reversibility of the excitation process itself and of the sample's initial state, limiting the application of pump-probe techniques to the investigation of fully reversible ultrafast dynamics.

We recently demonstrated a novel experimental approach, which overcomes these restrictions by continuous probing of a relaxation process with a single X-ray pulse \cite{buzzi_single-shot_2017}. For this we employ an off-axis Fresnel zone plate to stretch an incoming X-ray pulse while introducing at the same time an angular encoding of the arrival time of the X-rays. In our previous work, we have performed such an X-ray streaking experiment in a reflection geometry and used the resonant transverse magneto optic Kerr effect (T-MOKE) to follow the ultrafast magnetization dynamics in a thin transition metal film \cite{buzzi_single-shot_2017}. 
Since resonant T-MOKE provides a very strong magnetic dichroism ($40\%$ in our case \cite{buzzi_single-shot_2017}), it was ideally suited for a first feasibility demonstration. On the other hand, this experimental geometry can only be employed to study quantitatively the magnetization dynamics, and is limited to probing surface phenomena in materials exhibiting an in-plane magnetization. 

To demonstrate a broader applicability of this X-ray streaking technique we are reporting here a further development where an off-axis zone plate is used to generate a tilted wave front as a probe for X-ray absortion spectroscopy. Specifically, we use the X-ray Magnetic Circular Dichroism (XMCD) effect \cite{chen_experimental_1995, valencia_faraday_2006} in transmission to follow the magnetization dynamics in thin CoDy alloy films exhibiting out-of-plane magnetization. Our results demonstrate that signal variations of a few percent level can be resolved. It suggests that an absorption spectroscopy based X-ray streaking is suitable for the investigation of ultrafast dynamics in a variety of other fields \cite{ramasesha_real-time_2016, nguyen_study_2016, vaida_nonmetal_2018}.

Since the discovery of laser-induced ultrafast demagnetization by Beaurepaire et al \cite{beaurepaire_ultrafast_1996}, intense research to understand what is driving this phenomenon have been performed, leading to an other discovery of technological relevance: the possibility of single shot all optical switching in transition metal (TM) - rare earth (RE) alloys \cite{stanciu_all-optical_2007}. In order to better understand what is the influence of the RE elements on the dynamics of the TM, we have compared the magnetization dynamics of two CoDy alloys with different chemical compositions. In this study, we find that for higher Dy concentration the initial quenching observed at the Co site is slower, but reaches a higher degree of demagnetization. Furthermore, for each alloys, varying the pump fluence does not affect the time scale of the initial magnetization quenching.

\subsection{Experiment}
\begin{center}
\begin{figure}
 \includegraphics[width=0.49\textwidth]{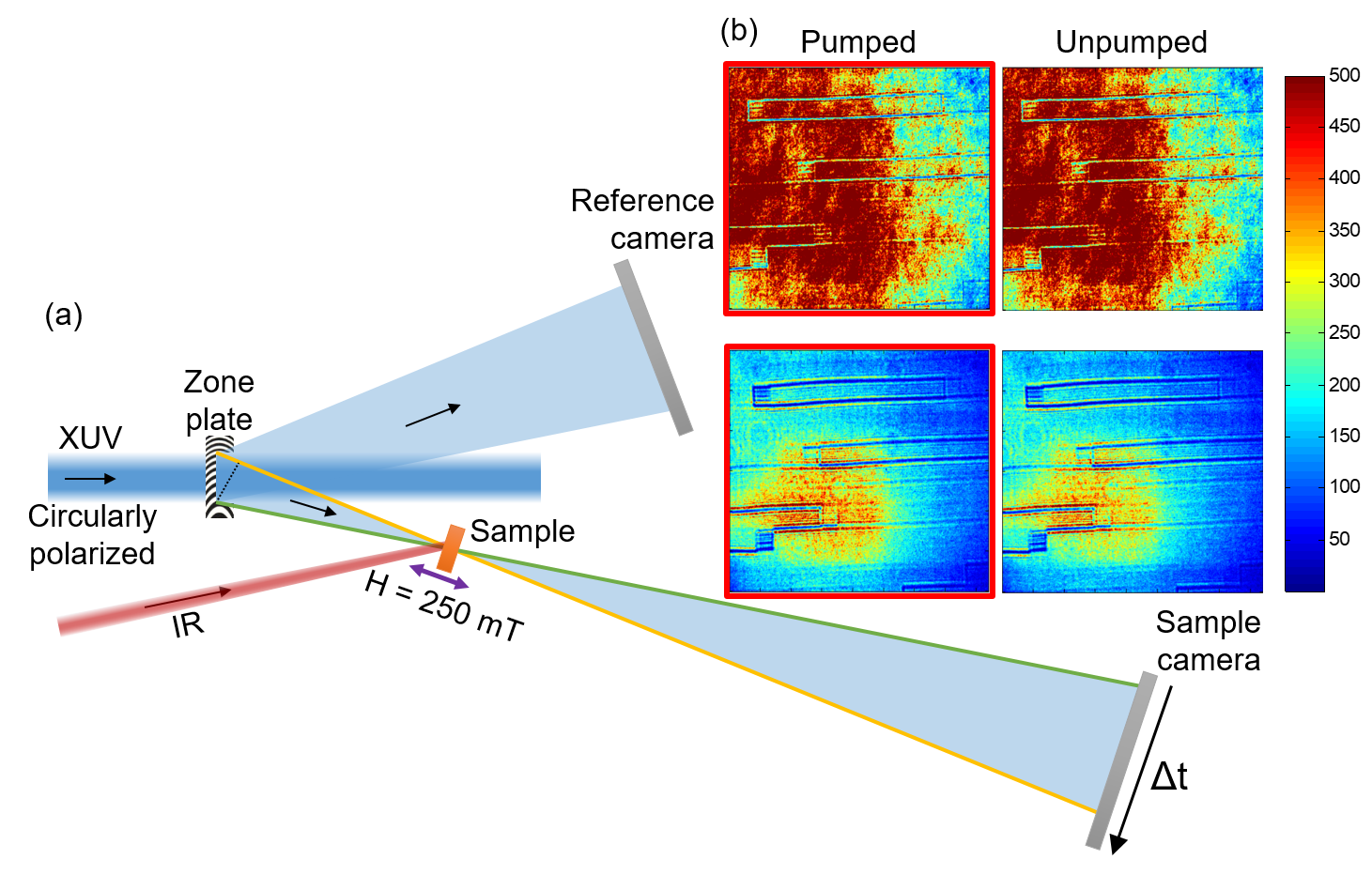}%
 \caption{(a) Sketch of the experimental setup. (b) Intensity distribution of a single X-ray pulse recorded by the \textit{sample camera} (converging first order, after passing through the Co$_{76}$Dy$_{24}$ thin film sample) and \textit{reference camera} (diverging first order) with (pumped) and without (unpumped) IR laser excitation. 
 \label{fig1}}
 \end{figure}
\end{center}

The experiment has been realized at the DiProI end station \cite{capotondi_invited_2013} of the seeded XUV-FEL FERMI at Elettra, Trieste \cite{allaria_highly_2012}, which provides femtosecond short X-ray pulses with full polarization control. The experimental layout of our zone plate based X-ray streaking experiment is shown in Fig.~\ref{fig1} (a). The off-axis zone plate is the same as used in our previous experiment \cite{buzzi_single-shot_2017}. To overfill its $4.84$ mm square aperture, the beamline focus was moved far behind the sample position by releasing the bend of the KB mirrors, producing a nearly collimated beam. The converging positive first order of the zone plate passes through the sample and its intensity distribution is collected by a CCD camera (so-called \textit{sample-camera} afterwards). Note that the photons detected at the bottom end of the \textit{sample-camera} (yellow line in Fig.~\ref{fig1} (a)) have traveled a longer path than those arriving at the top of the camera (green line in Fig.~\ref{fig1} (a)). Their arrival time difference at the zone plate focus ($\Delta t = 1.57$ ps) is proportional to the product of the number of zone pairs (23,000) and the X-ray wavelength (20.5 nm). The time evolution of the measured signal is therefore given by the intensity variations between those two points.  

The sample is positioned in the vicinity of the focus of the positive first diffraction order. This allows for controlling the X-ray fluence on the sample to avoid X-ray induced modifications, while limiting the different optical path, i.e. time delays, to probe different sample areas. A precise calibration of the time axis is obtained by varying the delay between IR and X-ray pulses. A permanent magnetic field of $\pm 250$ mT is applied to reset a single domain state after each pump-probe event. 

To characterize the shot to shot variations of the spatial intensity distribution of incident X-ray pulse, a second CCD camera (\textit{reference-camera}) is used to record the diverging negative first order. For a single X-ray pulse, two images are therefore recorded simultaneously: the signal characterizing the observed dynamics is encoded in the intensity variation of the \textit{sample-camera} image; and the normalization with respect to the incident intensity distribution is provided by the simultaneously recorded image of the \textit{reference-camera}. The higher diffraction orders of the diffractive element are spatially filtered along the beam path from the zone plate to the detector in order to minimize their contamination at the sensor plane.The respective zone plate to CCD camera distances have been chosen such that the pixel size does not limit the achievable time resolution and that comparable magnifications are obtained for both cameras, which simplifies data analysis. Aluminum filters have been placed in front of both cameras to block the IR transmitted/scattered after the sample. In order to get the same XUV contrast on both camera, the Al filter in front of the \textit{reference-camera} is much thicker than the one in front of the \textit{sample-camera}.

\begin{center}
\begin{figure}
 \includegraphics[width=0.45\textwidth]{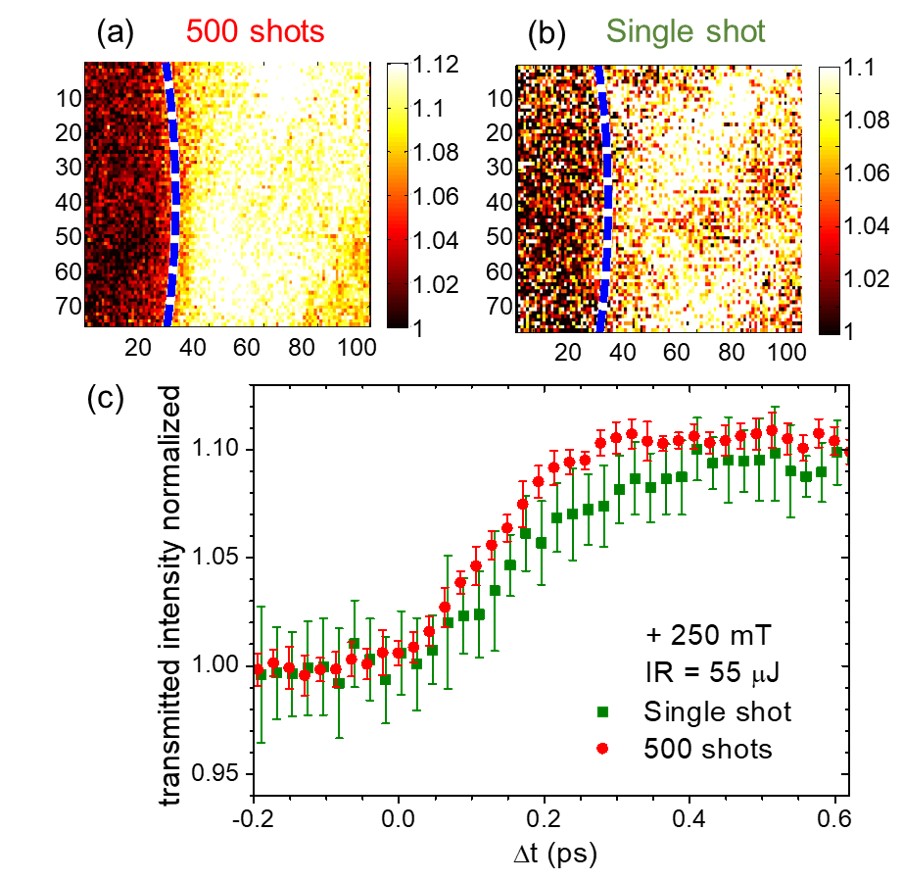}%
 \caption{Treated transmission signal image (see text) revealing the IR-pulse-induced magnetization dynamics of a thin Co$_{76}$Dy$_{24}$ film in the presence of an out-of-plane applied field of 250 mT and an IR pump pulse intensity of 55 $\micro$J. (a) The signal of 500 pump-probe cycles was accumulated to obtain a high signal-to-noise image. The dashed blue line indicates pump-probe time overlap. (b) Signal image obtained from a single pump-probe event. (c) Azimuthally integrated radial intensity of the signal images projected onto the time delay axis (red for 500 shots, green for single shot) normalized by the unpumped intensity. 
 \label{fig2}}
 \end{figure}
\end{center}

The images in Fig.~\ref{fig1} (b) show two pairs of single X-ray pulse images recorded by the \textit{reference-} (top line) and \textit{sample-camera} (bottom line). The two images in the left column were recorded with IR pulse excitation (pumped), while no excitation was applied in case of the right column (unpumped). Comparing the pumped and unpumped raw images, there are no clear evidences of a pump pulse induced evolution of the magnetization. However, as demonstrated in the next section, starting from these four images an appropriate normalization procedure yields an image that clearly reveals the signature of the laser-pulse-induced ultrafast magnetization dynamics (Fig.~\ref{fig2}).

The measurements reported in this article have been performed on 50 nm thick Co$_{1-x}$Dy$_x$ alloy films, which were deposited by magnetron sputtering on 30 nm thin Si$_3$N$_4$ membranes. A seed layer of 3 nm of Ta and 5 nm of Pt was used and the film was capped with a 5 nm thick Pt layer to prevent oxidation. Two different alloy compositions were investigated: Co$_{86}$Dy$_{14}$ and Co$_{76}$Dy$_{24}$. At room temperature, both films exhibit a magnetic out-of-plane anisotropy and a square hysteresis loop with a coercive field of 100 mT for Co$_{86}$Dy$_{14}$ and 200 mT for Co$_{76}$Dy$_{24}$. 

To realize a jitter free IR-pump/XUV-probe measurements, a fraction of FERMI's IR seed laser was used to excite the sample. As shown previously \cite{capotondi_invited_2013}, this yields sub 10 fs jitter between IR- and XUV pulse arrival time and repetitive, accumulative measurements can be realized without significant degradation of the overall time resolution. The wavelength of the linearly polarized IR pulses was centered at 780 nm and the pulse length is about 100 fs. The IR laser spot size at the sample position was measured, thanks to phosphor paint, to be about 500 x 400 $\micro$m$^2$ (FWHM). To reach the typical optical intensity range inducing sizable effects on magnetic response an IR pulse from 5.5 to 55 $\micro$J was used. 

The photon energy of the circularly polarized XUV pulses was tuned to 20.5 nm to match the magnetically dichroic M$_{2,3}$ absorption edge of Co (see note \cite{note}). Using the 12-th harmonic of the seed laser, the XUV pulse length is expected to be about 70 fs \cite{finetti_pulse_2017}. The XUV spot size in the sample plane was measured to be $\approx$ 150 x 100 $\micro m^2$ (FWHM), thus probing a homogeneous part of the IR pumped area. The XUV intensity at the sample plane was set far below the excitation regime \cite{wang_femtosecond_2012}.

\subsection{Results}
The as-recorded images shown in Fig.~\ref{fig1} exhibit very strong intensity fluctuations, which outweigh the signal variations from the IR-excitation-induced magnetization dynamics. These fluctuations have three origins: the spatially inhomogeneous intensity distribution of the incident XUV pulse; inhomogeneities in the XUV diffraction of the zone plate; and the thickness variations of the Al filter protecting the CCD cameras against the IR photons of the pump pulse. These filter thickness variations, in particular for the thicker Al filter of the \textit{reference-camera}, introduce substantial differences between the pattern recorded by \textit{sample} and \textit{reference cameras}. In addition, the zone plate diffracted images show artifacts due to stitching errors from the zone plate fabrication process, which give rise to phase contrast in the far field plane of both CCD cameras.

To extract the signature of the magnetization dynamics from the pumped image, we have to employ background images recorded by the two cameras without XUV probe pulse, and with and without IR pump pulse. These images have to be recorded each time there is any experimental change in parameters. A complete data set is thus composed of eight images, the four shown in Fig.~\ref{fig1} and four background pictures without XUV probe pulse. The applied procedure to extract the desired signal consists of first background correcting each XUV camera image and then normalizing for each camera the pumped by the corresponding unpumped image. This normalized image of the \textit{sample camera} is then divided by the one of the \textit{reference camera} to get the final signal map, whose intensity variation exhibits the IR-pulse-induced magnetization dynamics.

Final signal maps obtained by this procedure are shown in Fig.~\ref{fig2}. These are recorded by transmission of the XUV probe pulse through a thin Co$_{76}$Dy$_{24}$ film under the presence of an out-of-plane magnetic field of 250 mT. The IR pump pulse intensity was 55 $\mu$J. The image in Fig.~\ref{fig2} (a) is obtained by accumulating the signal for 500 pump-probe cycles, while a single pump-probe cycle yielded the image shown in Fig.~\ref{fig2} (b). The pump-probe delay axis proceeds mainly in the horizontal direction from left to right. The IR-pulse-induced magnetization dynamics is clearly visible in both images as an abrupt color change, which sets in right after temporal overlap between IR pump and XUV probe pulses (dashed blue line). 

X-rays diffracted from the same area of the zone plate probe the same delay thus their signal can be averaged using an angular integration (e.g., along the dashed blue line for zero delay). This yields the projection of the signal onto the pump-probe time delay axis shown in Fig.~\ref{fig2} (c) for the accumulation of 500 shots (red) and the single shot measurement (green). These intensities are normalized by the unpumped signal and are therefore equal to 1 before time 0. The good signal to noise ratio of the single shot measurement underlines the potential of this technique. In this particular configuration of circular polarization and magnetic field orientation, the laser-induced demagnetization leads to a signal increase reflecting the expected rapid quenching of the magnetization.

The two delay curves exhibit overall identical shapes, which underlines the reversibility of the laser-induced demagnetization process. The slightly lower degree of demagnetization observed in the single shot measurement suggests a lower IR pump fluence. This may be caused by shot-to-shot fluctuations of the pump laser intensity, or by pointing fluctuations of the XUV probe pulse altering the probed sample area with respect to the IR pump pulse. Note that single shot measurements overcome these limitations, emphasizing the relevance of this technique even when studying reversible processes.

\begin{center}
\begin{figure}
 \includegraphics[width=0.45\textwidth]{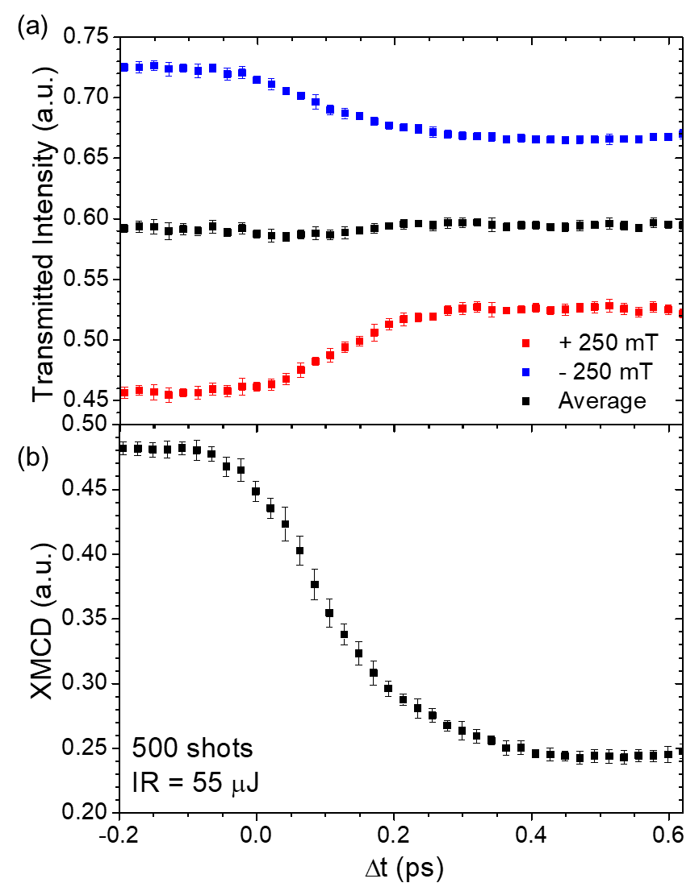}%
 \caption{(a) Non normalized transmitted intensity as a function of pump-probe time delay recorded on a Co$_{76}$Dy$_{24}$ film by averaging 500 pump-probe cycles (IR intensity of 55 $\mu$J). A static out-of-plane magnetic field of 250 mT was applied in positive (red) and negative (blue) direction. The black curve shows the arithmetic average of the two signals. (b) The logarithm of the ratio of the blue and red curves in (a) gives the temporal evolution of the XMCD signal.
 \label{fig3}}
 \end{figure}
\end{center}

Measuring magnetization dynamics for opposite direction of externally applied magnetic field allows to quantify the degree of the demagnetization dynamic. Figure \ref{fig3} (a) shows the result of such a measurement, non normalized by the unpumped signal, which has been performed on a Co$_{76}$Dy$_{24}$ film by averaging 500 pump-probe cycles and an IR pump pulse intensity of 55 $\mu$J. The blue and red curves show the temporal evolution of the transmitted XUV intensity for opposite out-of-plane field directions. The logarithmic of these curves give the absorption $\mu^{\pm}$, and the difference between the two absorptions, $\mu^-$ and $\mu^+$, gives the XMCD intensity (Fig.~\ref{fig3} (b)), which is directly proportional to the film's magnetization \cite{stohr_magnetism:_2006}. For negative delays, the separation between those two curves represents the unperturbed magnetization of the film's ground state, while the laser-induced ultrafast demagnetization leads to a reduction of this separation. In order to disentangle the electronic and magnetic dynamic of the film we use conventional analysis for dichroic response: where the average curve (shown by the black line in Fig.~\ref{fig3} (a)) takes into account for the electronic excitation while the logarithm of the ratio of those two curves collected with different orientation of the magnetic field, i.e. XMCD signal, represent the magnetization dynamic (Fig.~\ref{fig3} (b)). The nearly constant value of the average curve in Fig.~\ref{fig3} (a) shows that in this pumping regime the electronic response is almost negligible, while most of the sample response is due to the reduction of the XMCD signal. 
The laser-pulse-induced reduction of the XMCD amplitude indicates a demagnetization degree of about 50$\%$. 

Looking at the noise level of the single shot data in Fig.~\ref{fig2} (c) we can quantify our experimental detection limit to correspond to a 2$\%$ change in magnetization, which highlights the high sensitivity of this single shot technique. For multiple data exposure, the detection limit can be as low as 0.7$\%$ \cite{note3}.

\begin{center}
\begin{figure*}
 \includegraphics [width=0.98\textwidth]{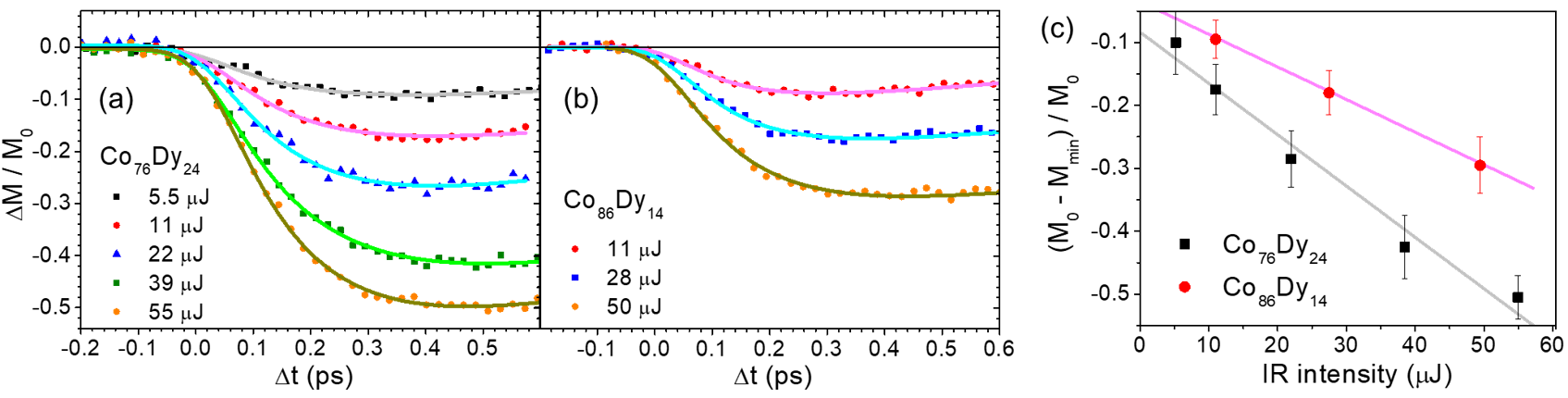}%
 \caption{Normalized demagnetization dynamics as a function of pump fluence recorded (a) on Co$_{76}$Dy$_{24}$ (500 shots accumulated) and (b) on Co$_{86}$Dy$_{14}$ (50 shots accumulated). (c) Degree of demagnetization as a function of the employed IR laser fluence for Co$_{76}$Dy$_{24}$ (black squares) and Co$_{86}$Dy$_{14}$ (red circles). The lines are a linear fit.
 \label{fig4}}
 \end{figure*}
\end{center}

The normalized demagnetization curves recorded on Co$_{76}$Dy$_{24}$ and Co$_{84}$Dy$_{16}$ for different IR pump fluence values are shown in Figure \ref{fig4}. To facilitate the comparison we plot the degree of demagnetization ($\Delta M / M_0$), which increases as expected with increasing pump fluence for each film. Comparing the results obtained for the two films we note that the film with higher Co concentration (Co$_{84}$Dy$_{16}$) exhibits a weaker demagnetization for equivalent IR pump intensity. Note that the relative IR pump - XUV probe alignment was verified in both cases.
The relationship between IR pump fluence and achieved degree of demagnetization is shown in Fig.~\ref{fig4} (c). The achieved degree of demagnetization increases for each film linearly with IR pump fluence over the sampled range of up to 50$\%$ of demagnetization, which is in agreement with previous studies on other Co containing TM-RE alloys \cite{alebrand_subpicosecond_2014, medapalli_efficiency_2012, beaulieu:tel-00953656}. The above mentioned lower degree of demagnetization for higher Co concentration is reflected by the different slopes.

To extract parameters from those demagnetization curves, we have fitted all data with a two exponential model \cite{malinowski_control_2008} (equation \ref{eq1}). 
\begin{equation}
\frac{dm}{dt}=\Gamma\left(t\right)\otimes \left\{a + b \Theta\left(t\right)\left[b + c \ast e^{-\frac{t}{\tau_m}}-d \ast e^{-\frac{t}{\tau_e}}\right]\right\}
\label{eq1}
\end{equation}
In this equation, $\Gamma\left(t\right)$ is the experimental time resolution, $\Theta\left(t\right)$ is the step function, $a$ represents the magnetization value before time zero ($t_0$), $\tau_m$ represents the characteristic demagnetization time and $\tau_e$ the characteristic remagnetization time.
Since the sampled time window extends only to 0.6 ps after $t_0$\cite{note2}, we have fixed the remagnetization parameter $\tau_e$ from time resolved MOKE measurements (not shown).

The experimental data are very well reproduced by these fits, which are shown by the lines in Fig.~\ref{fig4}. Within the uncertainty of this analysis, we find that for each film the demagnetization time does not depend on the employed IR pump fluence. The obtained values are $140 \pm 20$ fs for Co$_{76}$Dy$_{24}$ and $115 \pm 20$ fs for Co$_{86}$Dy$_{14}$, in good agreement with the work of Fert\'e \textit{et al} \cite{ferte_element-resolved_2017}.

\begin{center}
\begin{figure}
 \includegraphics[width=0.45\textwidth]{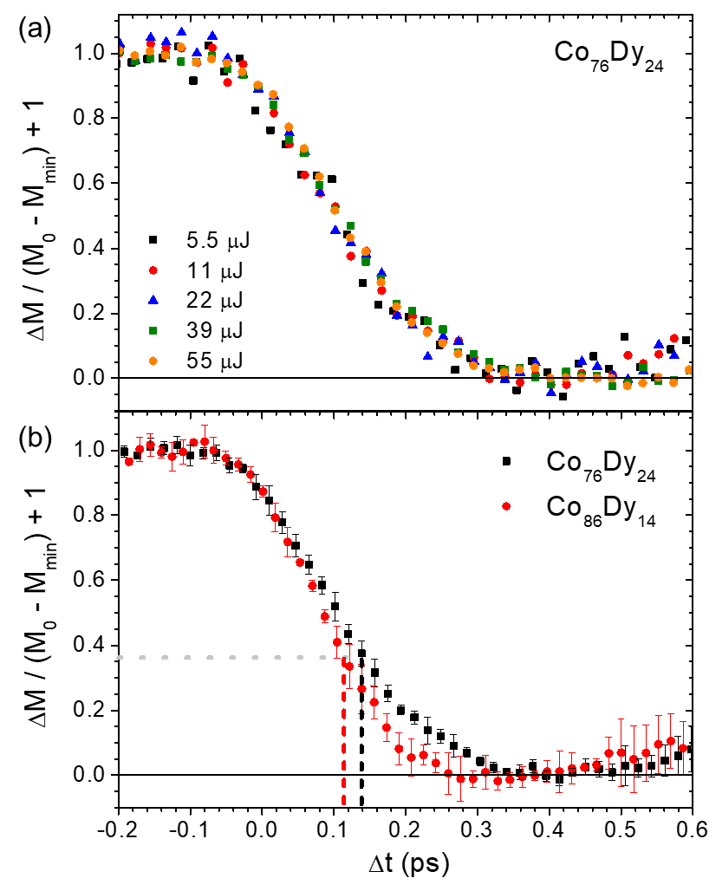}%
 \caption{(a) Scaled demagnetization curves recorded on Co$_{76}$Dy$_{24}$ for different IR fluence. (b) A comparison between the scaled curves of Co$_{76}$Dy$_{24}$ and Co$_{86}$Dy$_{14}$ (average of the respective fluence data) highlights the slower demagnetization time of the Dy rich alloy (Co$_{76}$Dy$_{24}$). 
\label{fig5}}
 \end{figure}
\end{center}

This numerical result can be confirmed visually by scaling the demagnetization curves, measured for different pump fluences, to equal amplitude as shown for Co$_{76}$Dy$_{24}$ in Fig.~\ref{fig5} (a). The absence of any systematic deviation between these data underlines that the IR pump fluence does not alter the observed demagnetization time. Figure \ref{fig5} (b) shows the average of those scaled curves for all the IR fluences, for Co$_{76}$Dy$_{24}$ (black) and Co$_{86}$Dy$_{14}$ (red), emphasizing the presence of a different demagnetization time scale for different chemical composition.

\subsection{Discussion}

Since the discovery of ultrafast magnetization dynamics \cite{beaurepaire_ultrafast_1996}, intense research efforts have been made to reveal the underlying mechanism driving the ultrafast demagnetization\cite{malinowski_control_2008, stamm_femtosecond_2007, boeglin_distinguishing_2010, koopmans_explaining_2010, battiato_superdiffusive_2010, radu_transient_2011, pfau_ultrafast_2012, vodungbo_laser-induced_2012,la-o-vorakiat_ultrafast_2012,rudolf_ultrafast_2012, graves_nanoscale_2013, bergeard_ultrafast_2014, wieczorek_separation_2015, jal_structural_2017, zhang_unifying_2017}. 
While these studies report overall very similar magnetization dynamics, one notices differences in details such as the specific characteristic time scale of magnetization quenching and partial recovery. These discrepancies may be sample structure related, but there are also indications that differences in probing depth of the employed techniques may play a role \cite{tengdin_critical_2018, you_revealing_2018}. It is in this regard important to note that in our transmission experiment the entire magnetic layer is probed. This implies that we are probing the average magnetization dynamics, regardless of the excitation profile.

Firstly, our data indicate that degree and time scale of the magnetization quenching depends on the chemical composition in CoDy alloys, thereby confirming previous results on TM-RE alloys. For example, Alebrand \textit{et al.}~\cite{alebrand_subpicosecond_2014} show that the efficiency of the magnetization quenching by optical excitation in the TM-TE ferrimagnets Tb$_x$Co$_{1-x}$ decreases with increasing TM content. This variation of the magnetization quenching with sample composition is related to the Curie temperature dependence on the alloy composition. From the study of Hansen \textit{et al.}~\cite{hansen_magnetic_1991} we know that the Curie temperature of Co$_{76}$Dy$_{24}$ is lower than the one of the Co$_{86}$Dy$_{14}$. Exciting both samples at room temperature, it means that for the same amount of deposited energy, the Co$_{76}$Dy$_{24}$ alloy reaches a temperature closer to the Curie temperature. 
Moreover, the dependence of the degree of magnetization quenching on the fluence can be related to the fact that for longer time delays, the remaining magnetization is proportional to the newly established equilibrium temperature \cite{Hohlfeld_nonequilibrium_1997}. Increasing the pump fluence increases this new equilibrium temperature and so the transient magnetization is smaller. The observation of a linear dependence indicates that over the employed fluence range the system remains far from the Curie temperature \cite{Roth_temperature_2012, Atxitia_evidence_2010}.

The lower Curie temperature of the Dy rich sample can also explain the observed slower demagnetization time of the TM in the CoDy ferrimagnet alloy. Indeed as pointed out by Suarez \textit{et al.} \cite{suarez_ultrafast_2015} and Atxitia \textit{et al.} \cite{Atxitia_controlling_2014} the complex demagnetization dynamic of ferrimagnetic films is strongly related to the concentration of the RE element that mediate the inter-sublattice exchange strength. In a phenomenological description, the numerical and analytical solutions of the Landau-Lifthitz-Bloch equation \cite{suarez_ultrafast_2015, Atxitia_controlling_2014} show that in such systems the demagnetization time of the TM element increases as a function of the RE concentration, assuming that the temperature remains sufficiently far from the Curie temperature. This consideration is in agreement with what we observe experimentally. In Fig.~\ref{fig5} (b) one notices that when probing the magnetization dynamics of the TM element a slower quenching is observed for the Dy rich alloy (Co$_{76}$Dy$_{24}$). We remark that the different compensation temperatures of our two CoDy alloys seem not to play a role here as demonstrated previously by Fert\'e \textit{et al} \cite{ferte_element-resolved_2017}.

Secondly, our results clearly demonstrate that for a given CoDy alloy the demagnetization time does not depend on the excitation fluence. This is in agreement with previous study where we are probing the magnetization dynamics with resonant magnetic small angle X-ray scattering \cite{vodungbo_laser-induced_2012, vodungbo_indirect_2016}. However, this finding is in contradiction with results from all-optical pump-probe experiments, which find a clear dependency of the demagnetization time on the pump fluence \cite{koopmans_explaining_2010, moisan_investigating_2014, mendil_resolving_2014, kuiper_spin-orbit_2014}. The main experimental difference between our x-ray experiments and the all-optical experiments is the respective sampling depth. While in all-optical experiments, the dynamics are usually probed in reflection, implying that the dynamics are sampled with exponentially decreasing sensitivity into the bulk of the film; the entire film is probed in our transmission geometry. Our findings may therefore support recent considerations that the probing depth of the employed techniques should be taken into consideration \cite{you_revealing_2018}. 

The observed fluence independency of the demagnetization time can be understood within the model put forward by Suarez \textit{et al.} \cite{suarez_ultrafast_2015} and Atxitia \textit{et al.} \cite{Atxitia_controlling_2014}. Their model predicts only a week fluence dependence of the demagnetization time as long as the system is not heated above the Curie temperature, which is true in our case, as discussed above. On the other hand, when the temperature reaches or overcomes the Curie temperature, the situation changes as reported by simulation \cite{suarez_ultrafast_2015, Atxitia_controlling_2014} and recent experimental work \cite{ferte_element-resolved_2017, you_revealing_2018}.

\subsection{Conclusions }
In conclusion, X-ray streaking in transmission geometry has been demonstrated by performing single shot based time resolved XMCD spectroscopy. This new experimental technique is sensitive to signal change as small as 2$\%$ and opens the path for the investigation of the dynamics of irreversible phenomena in a wide variety of fields. 

Our study on CoDy alloys reveals that when the average magnetization dynamics over the entire film thickness is probed the demagnetization time does not depend on the amount of energy deposited in the sample, but seems to be intrinsically given by the alloy composition.

Further development of the employed zone plate optic holds significant potential to further increase the capabilities of this novel technique. First of all, improving the manufacturing process will reduce the artifacts currently hampering the achievable signal quality. In addition, the fabrication of off-axis zone plates with finer structures will enable to extend the time window. Most importantly, more complex zone plate schemes can be envisioned to extend the method to probe, for example, several absorption edges simultaneously.

\subsection{Acknowledgments}
The authors are grateful for financial support received from the following agencies: The French “Agence National de la Recherche” via Projects No. UMAMI ANR-15-CE24-0009, the CNRS-PICS program, the CNRS-MOMENTUM program, and the SNSF project (number 200021\_160186). Access to Synchrotron SOLEIL through proposal ID 20160880 for characterization of static properties of the CoDy films is acknowledged.

\bibliography{ZP-CoDy_biblio}

\begin{thebibliography}{54}%
\makeatletter
\providecommand \@ifxundefined [1]{%
 \@ifx{#1\undefined}
}%
\providecommand \@ifnum [1]{%
 \ifnum #1\expandafter \@firstoftwo
 \else \expandafter \@secondoftwo
 \fi
}%
\providecommand \@ifx [1]{%
 \ifx #1\expandafter \@firstoftwo
 \else \expandafter \@secondoftwo
 \fi
}%
\providecommand \natexlab [1]{#1}%
\providecommand \enquote  [1]{``#1''}%
\providecommand \bibnamefont  [1]{#1}%
\providecommand \bibfnamefont [1]{#1}%
\providecommand \citenamefont [1]{#1}%
\providecommand \href@noop [0]{\@secondoftwo}%
\providecommand \href [0]{\begingroup \@sanitize@url \@href}%
\providecommand \@href[1]{\@@startlink{#1}\@@href}%
\providecommand \@@href[1]{\endgroup#1\@@endlink}%
\providecommand \@sanitize@url [0]{\catcode `\\12\catcode `\$12\catcode
  `\&12\catcode `\#12\catcode `\^12\catcode `\_12\catcode `\%12\relax}%
\providecommand \@@startlink[1]{}%
\providecommand \@@endlink[0]{}%
\providecommand \url  [0]{\begingroup\@sanitize@url \@url }%
\providecommand \@url [1]{\endgroup\@href {#1}{\urlprefix }}%
\providecommand \urlprefix  [0]{URL }%
\providecommand \Eprint [0]{\href }%
\providecommand \doibase [0]{http://dx.doi.org/}%
\providecommand \selectlanguage [0]{\@gobble}%
\providecommand \bibinfo  [0]{\@secondoftwo}%
\providecommand \bibfield  [0]{\@secondoftwo}%
\providecommand \translation [1]{[#1]}%
\providecommand \BibitemOpen [0]{}%
\providecommand \bibitemStop [0]{}%
\providecommand \bibitemNoStop [0]{.\EOS\space}%
\providecommand \EOS [0]{\spacefactor3000\relax}%
\providecommand \BibitemShut  [1]{\csname bibitem#1\endcsname}%
\let\auto@bib@innerbib\@empty
\bibitem [{\citenamefont {Squibb}\ \emph {et~al.}(2018)\citenamefont {Squibb},
  \citenamefont {Sapunar}, \citenamefont {Ponzi}, \citenamefont {Richter},
  \citenamefont {Kivim\"aki}, \citenamefont {Plekan}, \citenamefont {Finetti},
  \citenamefont {Sisourat}, \citenamefont {Zhaunerchyk}, \citenamefont
  {Marchenko}, \citenamefont {Journel}, \citenamefont {Guillemin},
  \citenamefont {Cucini}, \citenamefont {Coreno}, \citenamefont {Grazioli},
  \citenamefont {Di~Fraia}, \citenamefont {Callegari}, \citenamefont {Prince},
  \citenamefont {Decleva}, \citenamefont {Simon}, \citenamefont {Eland},
  \citenamefont {N.}, \citenamefont {Feifel},\ and\ \citenamefont
  {Piancastelli}}]{squibb_acetylacetone_2018}%
  \BibitemOpen
  \bibfield  {author} {\bibinfo {author} {\bibfnamefont {R.~J.}\ \bibnamefont
  {Squibb}}, \bibinfo {author} {\bibfnamefont {M.}~\bibnamefont {Sapunar}},
  \bibinfo {author} {\bibfnamefont {A.}~\bibnamefont {Ponzi}}, \bibinfo
  {author} {\bibfnamefont {R.}~\bibnamefont {Richter}}, \bibinfo {author}
  {\bibfnamefont {A.}~\bibnamefont {Kivim\"aki}}, \bibinfo {author}
  {\bibfnamefont {O.}~\bibnamefont {Plekan}}, \bibinfo {author} {\bibfnamefont
  {P.}~\bibnamefont {Finetti}}, \bibinfo {author} {\bibfnamefont
  {N.}~\bibnamefont {Sisourat}}, \bibinfo {author} {\bibfnamefont
  {V.}~\bibnamefont {Zhaunerchyk}}, \bibinfo {author} {\bibfnamefont
  {T.}~\bibnamefont {Marchenko}}, \bibinfo {author} {\bibfnamefont
  {L.}~\bibnamefont {Journel}}, \bibinfo {author} {\bibfnamefont
  {R.}~\bibnamefont {Guillemin}}, \bibinfo {author} {\bibfnamefont
  {R.}~\bibnamefont {Cucini}}, \bibinfo {author} {\bibfnamefont
  {M.}~\bibnamefont {Coreno}}, \bibinfo {author} {\bibfnamefont
  {C.}~\bibnamefont {Grazioli}}, \bibinfo {author} {\bibfnamefont
  {M.}~\bibnamefont {Di~Fraia}}, \bibinfo {author} {\bibfnamefont
  {C.}~\bibnamefont {Callegari}}, \bibinfo {author} {\bibfnamefont {K.~C.}\
  \bibnamefont {Prince}}, \bibinfo {author} {\bibfnamefont {P.}~\bibnamefont
  {Decleva}}, \bibinfo {author} {\bibfnamefont {M.}~\bibnamefont {Simon}},
  \bibinfo {author} {\bibfnamefont {J.~H.~D.}\ \bibnamefont {Eland}}, \bibinfo
  {author} {\bibfnamefont {D.}~\bibnamefont {N.}}, \bibinfo {author}
  {\bibfnamefont {R.}~\bibnamefont {Feifel}}, \ and\ \bibinfo {author}
  {\bibfnamefont {M.~N.}\ \bibnamefont {Piancastelli}},\ }\href {\doibase
  10.1038/s41467-017-02478-0} {\bibfield  {journal} {\bibinfo  {journal}
  {Nature Communications}\ }\textbf {\bibinfo {volume} {9}},\ \bibinfo {pages}
  {63} (\bibinfo {year} {2018})}\BibitemShut {NoStop}%
\bibitem [{\citenamefont {Rudenko}\ \emph {et~al.}(2017)\citenamefont
  {Rudenko}, \citenamefont {Inhester}, \citenamefont {Hanasaki}, \citenamefont
  {Li}, \citenamefont {Robatjazi}, \citenamefont {Erk}, \citenamefont {Boll},
  \citenamefont {Toyota}, \citenamefont {Hao}, \citenamefont {Vendrell},
  \citenamefont {Bomme}, \citenamefont {Savelyev}, \citenamefont {Rudek},
  \citenamefont {Foucar}, \citenamefont {Southworth}, \citenamefont {Lehmann},
  \citenamefont {Kraessig}, \citenamefont {Marchenko}, \citenamefont {Simon},
  \citenamefont {Ueda}, \citenamefont {Ferguson}, \citenamefont {Bucher},
  \citenamefont {Gorkhover}, \citenamefont {Carron}, \citenamefont
  {Alonso-Mori}, \citenamefont {Koglin}, \citenamefont {Correa}, \citenamefont
  {Williams}, \citenamefont {Boutet}, \citenamefont {Young}, \citenamefont
  {Bostedt}, \citenamefont {Son}, \citenamefont {Santra},\ and\ \citenamefont
  {Rolles}}]{rudenko_femtosecond_2017}%
  \BibitemOpen
  \bibfield  {author} {\bibinfo {author} {\bibfnamefont {A.}~\bibnamefont
  {Rudenko}}, \bibinfo {author} {\bibfnamefont {L.}~\bibnamefont {Inhester}},
  \bibinfo {author} {\bibfnamefont {K.}~\bibnamefont {Hanasaki}}, \bibinfo
  {author} {\bibfnamefont {X.}~\bibnamefont {Li}}, \bibinfo {author}
  {\bibfnamefont {S.~J.}\ \bibnamefont {Robatjazi}}, \bibinfo {author}
  {\bibfnamefont {B.}~\bibnamefont {Erk}}, \bibinfo {author} {\bibfnamefont
  {R.}~\bibnamefont {Boll}}, \bibinfo {author} {\bibfnamefont {K.}~\bibnamefont
  {Toyota}}, \bibinfo {author} {\bibfnamefont {Y.}~\bibnamefont {Hao}},
  \bibinfo {author} {\bibfnamefont {O.}~\bibnamefont {Vendrell}}, \bibinfo
  {author} {\bibfnamefont {C.}~\bibnamefont {Bomme}}, \bibinfo {author}
  {\bibfnamefont {E.}~\bibnamefont {Savelyev}}, \bibinfo {author}
  {\bibfnamefont {B.}~\bibnamefont {Rudek}}, \bibinfo {author} {\bibfnamefont
  {L.}~\bibnamefont {Foucar}}, \bibinfo {author} {\bibfnamefont {S.~H.}\
  \bibnamefont {Southworth}}, \bibinfo {author} {\bibfnamefont {C.~S.}\
  \bibnamefont {Lehmann}}, \bibinfo {author} {\bibfnamefont {B.}~\bibnamefont
  {Kraessig}}, \bibinfo {author} {\bibfnamefont {T.}~\bibnamefont {Marchenko}},
  \bibinfo {author} {\bibfnamefont {M.}~\bibnamefont {Simon}}, \bibinfo
  {author} {\bibfnamefont {K.}~\bibnamefont {Ueda}}, \bibinfo {author}
  {\bibfnamefont {K.~R.}\ \bibnamefont {Ferguson}}, \bibinfo {author}
  {\bibfnamefont {M.}~\bibnamefont {Bucher}}, \bibinfo {author} {\bibfnamefont
  {T.}~\bibnamefont {Gorkhover}}, \bibinfo {author} {\bibfnamefont
  {S.}~\bibnamefont {Carron}}, \bibinfo {author} {\bibfnamefont
  {R.}~\bibnamefont {Alonso-Mori}}, \bibinfo {author} {\bibfnamefont {J.~E.}\
  \bibnamefont {Koglin}}, \bibinfo {author} {\bibfnamefont {J.}~\bibnamefont
  {Correa}}, \bibinfo {author} {\bibfnamefont {G.~J.}\ \bibnamefont
  {Williams}}, \bibinfo {author} {\bibfnamefont {S.}~\bibnamefont {Boutet}},
  \bibinfo {author} {\bibfnamefont {L.}~\bibnamefont {Young}}, \bibinfo
  {author} {\bibfnamefont {C.}~\bibnamefont {Bostedt}}, \bibinfo {author}
  {\bibfnamefont {S.-K.}\ \bibnamefont {Son}}, \bibinfo {author} {\bibfnamefont
  {R.}~\bibnamefont {Santra}}, \ and\ \bibinfo {author} {\bibfnamefont
  {D.}~\bibnamefont {Rolles}},\ }\href {http://dx.doi.org/10.1038/nature22373}
  {\bibfield  {journal} {\bibinfo  {journal} {Nature}\ }\textbf {\bibinfo
  {volume} {546}},\ \bibinfo {pages} {129} (\bibinfo {year}
  {2017})}\BibitemShut {NoStop}%
\bibitem [{\citenamefont {Young}\ \emph {et~al.}(2010)\citenamefont {Young},
  \citenamefont {Kanter}, \citenamefont {Kr\"assig}, \citenamefont {Li},
  \citenamefont {March}, \citenamefont {Pratt}, \citenamefont {Santra},
  \citenamefont {Southworth}, \citenamefont {Rohringer}, \citenamefont
  {DiMauro}, \citenamefont {Doumy}, \citenamefont {Roedig}, \citenamefont
  {Berrah}, \citenamefont {Fang}, \citenamefont {Hoener}, \citenamefont
  {Bucksbaum}, \citenamefont {Cryan}, \citenamefont {Ghimire}, \citenamefont
  {Glownia}, \citenamefont {Reis}, \citenamefont {Bozek}, \citenamefont
  {Bostedt},\ and\ \citenamefont {Messerschmidt}}]{young_femtosecond_2010}%
  \BibitemOpen
  \bibfield  {author} {\bibinfo {author} {\bibfnamefont {L.}~\bibnamefont
  {Young}}, \bibinfo {author} {\bibfnamefont {E.~P.}\ \bibnamefont {Kanter}},
  \bibinfo {author} {\bibfnamefont {B.}~\bibnamefont {Kr\"assig}}, \bibinfo
  {author} {\bibfnamefont {Y.}~\bibnamefont {Li}}, \bibinfo {author}
  {\bibfnamefont {A.~M.}\ \bibnamefont {March}}, \bibinfo {author}
  {\bibfnamefont {S.~T.}\ \bibnamefont {Pratt}}, \bibinfo {author}
  {\bibfnamefont {R.}~\bibnamefont {Santra}}, \bibinfo {author} {\bibfnamefont
  {S.~H.}\ \bibnamefont {Southworth}}, \bibinfo {author} {\bibfnamefont
  {N.}~\bibnamefont {Rohringer}}, \bibinfo {author} {\bibfnamefont {L.~F.}\
  \bibnamefont {DiMauro}}, \bibinfo {author} {\bibfnamefont {G.}~\bibnamefont
  {Doumy}}, \bibinfo {author} {\bibfnamefont {C.~A.}\ \bibnamefont {Roedig}},
  \bibinfo {author} {\bibfnamefont {N.}~\bibnamefont {Berrah}}, \bibinfo
  {author} {\bibfnamefont {L.}~\bibnamefont {Fang}}, \bibinfo {author}
  {\bibfnamefont {M.}~\bibnamefont {Hoener}}, \bibinfo {author} {\bibfnamefont
  {P.~H.}\ \bibnamefont {Bucksbaum}}, \bibinfo {author} {\bibfnamefont {J.~P.}\
  \bibnamefont {Cryan}}, \bibinfo {author} {\bibfnamefont {S.}~\bibnamefont
  {Ghimire}}, \bibinfo {author} {\bibfnamefont {J.~M.}\ \bibnamefont
  {Glownia}}, \bibinfo {author} {\bibfnamefont {D.~A.}\ \bibnamefont {Reis}},
  \bibinfo {author} {\bibfnamefont {J.~D.}\ \bibnamefont {Bozek}}, \bibinfo
  {author} {\bibfnamefont {C.}~\bibnamefont {Bostedt}}, \ and\ \bibinfo
  {author} {\bibfnamefont {M.}~\bibnamefont {Messerschmidt}},\ }\href {\doibase
  10.1038/nature09177} {\bibfield  {journal} {\bibinfo  {journal} {Nature}\
  }\textbf {\bibinfo {volume} {466}},\ \bibinfo {pages} {56} (\bibinfo {year}
  {2010})}\BibitemShut {NoStop}%
\bibitem [{\citenamefont {Coquelle}\ \emph {et~al.}(2017)\citenamefont
  {Coquelle}, \citenamefont {Sliwa}, \citenamefont {Woodhouse}, \citenamefont
  {Schir\'o}, \citenamefont {Adam}, \citenamefont {Aquila}, \citenamefont
  {Barends}, \citenamefont {Boutet}, \citenamefont {Byrdin}, \citenamefont
  {Carbajo}, \citenamefont {De~la Mora}, \citenamefont {Doak}, \citenamefont
  {Feliks}, \citenamefont {Fieschi}, \citenamefont {Foucar}, \citenamefont
  {Guillon}, \citenamefont {Hilpert}, \citenamefont {Hunter}, \citenamefont
  {Jakobs}, \citenamefont {Koglin}, \citenamefont {Kovacsova}, \citenamefont
  {Lane}, \citenamefont {L\'evy}, \citenamefont {Liang}, \citenamefont {Nass},
  \citenamefont {Ridard}, \citenamefont {Robinson}, \citenamefont {Roome},
  \citenamefont {Ruckebusch}, \citenamefont {Seaberg}, \citenamefont {Thepaut},
  \citenamefont {Cammarata}, \citenamefont {Demachy}, \citenamefont {Field},
  \citenamefont {Shoeman}, \citenamefont {Bourgeois}, \citenamefont
  {Colletier}, \citenamefont {Schlichting},\ and\ \citenamefont
  {Weik}}]{coquelle_chromophore_2017}%
  \BibitemOpen
  \bibfield  {author} {\bibinfo {author} {\bibfnamefont {N.}~\bibnamefont
  {Coquelle}}, \bibinfo {author} {\bibfnamefont {M.}~\bibnamefont {Sliwa}},
  \bibinfo {author} {\bibfnamefont {J.}~\bibnamefont {Woodhouse}}, \bibinfo
  {author} {\bibfnamefont {G.}~\bibnamefont {Schir\'o}}, \bibinfo {author}
  {\bibfnamefont {V.}~\bibnamefont {Adam}}, \bibinfo {author} {\bibfnamefont
  {A.}~\bibnamefont {Aquila}}, \bibinfo {author} {\bibfnamefont {T.~R.~M.}\
  \bibnamefont {Barends}}, \bibinfo {author} {\bibfnamefont {S.}~\bibnamefont
  {Boutet}}, \bibinfo {author} {\bibfnamefont {M.}~\bibnamefont {Byrdin}},
  \bibinfo {author} {\bibfnamefont {S.}~\bibnamefont {Carbajo}}, \bibinfo
  {author} {\bibfnamefont {E.}~\bibnamefont {De~la Mora}}, \bibinfo {author}
  {\bibfnamefont {R.~B.}\ \bibnamefont {Doak}}, \bibinfo {author}
  {\bibfnamefont {M.}~\bibnamefont {Feliks}}, \bibinfo {author} {\bibfnamefont
  {F.}~\bibnamefont {Fieschi}}, \bibinfo {author} {\bibfnamefont
  {L.}~\bibnamefont {Foucar}}, \bibinfo {author} {\bibfnamefont
  {V.}~\bibnamefont {Guillon}}, \bibinfo {author} {\bibfnamefont
  {M.}~\bibnamefont {Hilpert}}, \bibinfo {author} {\bibfnamefont {M.~S.}\
  \bibnamefont {Hunter}}, \bibinfo {author} {\bibfnamefont {S.}~\bibnamefont
  {Jakobs}}, \bibinfo {author} {\bibfnamefont {J.~E.}\ \bibnamefont {Koglin}},
  \bibinfo {author} {\bibfnamefont {G.}~\bibnamefont {Kovacsova}}, \bibinfo
  {author} {\bibfnamefont {T.~J.}\ \bibnamefont {Lane}}, \bibinfo {author}
  {\bibfnamefont {B.}~\bibnamefont {L\'evy}}, \bibinfo {author} {\bibfnamefont
  {M.}~\bibnamefont {Liang}}, \bibinfo {author} {\bibfnamefont
  {K.}~\bibnamefont {Nass}}, \bibinfo {author} {\bibfnamefont {J.}~\bibnamefont
  {Ridard}}, \bibinfo {author} {\bibfnamefont {J.~S.}\ \bibnamefont
  {Robinson}}, \bibinfo {author} {\bibfnamefont {C.~M.}\ \bibnamefont {Roome}},
  \bibinfo {author} {\bibfnamefont {C.}~\bibnamefont {Ruckebusch}}, \bibinfo
  {author} {\bibfnamefont {M.}~\bibnamefont {Seaberg}}, \bibinfo {author}
  {\bibfnamefont {M.}~\bibnamefont {Thepaut}}, \bibinfo {author} {\bibfnamefont
  {M.}~\bibnamefont {Cammarata}}, \bibinfo {author} {\bibfnamefont
  {I.}~\bibnamefont {Demachy}}, \bibinfo {author} {\bibfnamefont
  {M.}~\bibnamefont {Field}}, \bibinfo {author} {\bibfnamefont {R.~L.}\
  \bibnamefont {Shoeman}}, \bibinfo {author} {\bibfnamefont {D.}~\bibnamefont
  {Bourgeois}}, \bibinfo {author} {\bibfnamefont {J.-P.}\ \bibnamefont
  {Colletier}}, \bibinfo {author} {\bibfnamefont {I.}~\bibnamefont
  {Schlichting}}, \ and\ \bibinfo {author} {\bibfnamefont {M.}~\bibnamefont
  {Weik}},\ }\href {http://dx.doi.org/10.1038/nchem.2853} {\bibfield  {journal}
  {\bibinfo  {journal} {Nature Chemistry}\ }\textbf {\bibinfo {volume} {10}},\
  \bibinfo {pages} {31} (\bibinfo {year} {2017})}\BibitemShut {NoStop}%
\bibitem [{\citenamefont {Tayeb-Fligelman}\ \emph {et~al.}(2017)\citenamefont
  {Tayeb-Fligelman}, \citenamefont {Tabachnikov}, \citenamefont {Moshe},
  \citenamefont {Goldshmidt-Tran}, \citenamefont {Sawaya}, \citenamefont
  {Coquelle}, \citenamefont {Colletier},\ and\ \citenamefont
  {Landau}}]{tayeb-fligelman_cytotoxic_2017}%
  \BibitemOpen
  \bibfield  {author} {\bibinfo {author} {\bibfnamefont {E.}~\bibnamefont
  {Tayeb-Fligelman}}, \bibinfo {author} {\bibfnamefont {O.}~\bibnamefont
  {Tabachnikov}}, \bibinfo {author} {\bibfnamefont {A.}~\bibnamefont {Moshe}},
  \bibinfo {author} {\bibfnamefont {O.}~\bibnamefont {Goldshmidt-Tran}},
  \bibinfo {author} {\bibfnamefont {M.~R.}\ \bibnamefont {Sawaya}}, \bibinfo
  {author} {\bibfnamefont {N.}~\bibnamefont {Coquelle}}, \bibinfo {author}
  {\bibfnamefont {J.-P.}\ \bibnamefont {Colletier}}, \ and\ \bibinfo {author}
  {\bibfnamefont {M.}~\bibnamefont {Landau}},\ }\href {\doibase
  10.1126/science.aaf4901} {\bibfield  {journal} {\bibinfo  {journal}
  {Science}\ }\textbf {\bibinfo {volume} {355}},\ \bibinfo {pages} {831}
  (\bibinfo {year} {2017})}\BibitemShut {NoStop}%
\bibitem [{\citenamefont {Graves}\ \emph {et~al.}(2013)\citenamefont {Graves},
  \citenamefont {Reid}, \citenamefont {Wang}, \citenamefont {Wu}, \citenamefont
  {de~Jong}, \citenamefont {Vahaplar}, \citenamefont {Radu}, \citenamefont
  {Bernstein}, \citenamefont {Messerschmidt}, \citenamefont {M\"uller},
  \citenamefont {Coffee}, \citenamefont {Bionta}, \citenamefont {Epp},
  \citenamefont {Hartmann}, \citenamefont {Kimmel}, \citenamefont {Hauser},
  \citenamefont {Hartmann}, \citenamefont {Holl}, \citenamefont {Gorke},
  \citenamefont {Mentink}, \citenamefont {Tsukamoto}, \citenamefont {Fognini},
  \citenamefont {Turner}, \citenamefont {Schlotter}, \citenamefont {Rolles},
  \citenamefont {Soltau}, \citenamefont {Str\"uder}, \citenamefont {Acremann},
  \citenamefont {Kimel}, \citenamefont {Kirilyuk}, \citenamefont {Rasing},
  \citenamefont {St\"ohr}, \citenamefont {Scherz},\ and\ \citenamefont
  {D\"urr}}]{graves_nanoscale_2013}%
  \BibitemOpen
  \bibfield  {author} {\bibinfo {author} {\bibfnamefont {C.~E.}\ \bibnamefont
  {Graves}}, \bibinfo {author} {\bibfnamefont {A.~H.}\ \bibnamefont {Reid}},
  \bibinfo {author} {\bibfnamefont {T.}~\bibnamefont {Wang}}, \bibinfo {author}
  {\bibfnamefont {B.}~\bibnamefont {Wu}}, \bibinfo {author} {\bibfnamefont
  {S.}~\bibnamefont {de~Jong}}, \bibinfo {author} {\bibfnamefont
  {K.}~\bibnamefont {Vahaplar}}, \bibinfo {author} {\bibfnamefont
  {I.}~\bibnamefont {Radu}}, \bibinfo {author} {\bibfnamefont {D.~P.}\
  \bibnamefont {Bernstein}}, \bibinfo {author} {\bibfnamefont {M.}~\bibnamefont
  {Messerschmidt}}, \bibinfo {author} {\bibfnamefont {L.}~\bibnamefont
  {M\"uller}}, \bibinfo {author} {\bibfnamefont {R.}~\bibnamefont {Coffee}},
  \bibinfo {author} {\bibfnamefont {M.}~\bibnamefont {Bionta}}, \bibinfo
  {author} {\bibfnamefont {S.~W.}\ \bibnamefont {Epp}}, \bibinfo {author}
  {\bibfnamefont {R.}~\bibnamefont {Hartmann}}, \bibinfo {author}
  {\bibfnamefont {N.}~\bibnamefont {Kimmel}}, \bibinfo {author} {\bibfnamefont
  {G.}~\bibnamefont {Hauser}}, \bibinfo {author} {\bibfnamefont
  {A.}~\bibnamefont {Hartmann}}, \bibinfo {author} {\bibfnamefont
  {P.}~\bibnamefont {Holl}}, \bibinfo {author} {\bibfnamefont {H.}~\bibnamefont
  {Gorke}}, \bibinfo {author} {\bibfnamefont {J.~H.}\ \bibnamefont {Mentink}},
  \bibinfo {author} {\bibfnamefont {A.}~\bibnamefont {Tsukamoto}}, \bibinfo
  {author} {\bibfnamefont {A.}~\bibnamefont {Fognini}}, \bibinfo {author}
  {\bibfnamefont {J.~J.}\ \bibnamefont {Turner}}, \bibinfo {author}
  {\bibfnamefont {W.~F.}\ \bibnamefont {Schlotter}}, \bibinfo {author}
  {\bibfnamefont {D.}~\bibnamefont {Rolles}}, \bibinfo {author} {\bibfnamefont
  {H.}~\bibnamefont {Soltau}}, \bibinfo {author} {\bibfnamefont
  {L.}~\bibnamefont {Str\"uder}}, \bibinfo {author} {\bibfnamefont
  {Y.}~\bibnamefont {Acremann}}, \bibinfo {author} {\bibfnamefont {A.~V.}\
  \bibnamefont {Kimel}}, \bibinfo {author} {\bibfnamefont {A.}~\bibnamefont
  {Kirilyuk}}, \bibinfo {author} {\bibfnamefont {T.}~\bibnamefont {Rasing}},
  \bibinfo {author} {\bibfnamefont {J.}~\bibnamefont {St\"ohr}}, \bibinfo
  {author} {\bibfnamefont {A.~O.}\ \bibnamefont {Scherz}}, \ and\ \bibinfo
  {author} {\bibfnamefont {H.~A.}\ \bibnamefont {D\"urr}},\ }\href {\doibase
  10.1038/nmat3597} {\bibfield  {journal} {\bibinfo  {journal} {Nature
  Materials}\ }\textbf {\bibinfo {volume} {12}},\ \bibinfo {pages} {293}
  (\bibinfo {year} {2013})}\BibitemShut {NoStop}%
\bibitem [{\citenamefont {Radu}\ \emph {et~al.}(2011)\citenamefont {Radu},
  \citenamefont {Vahaplar}, \citenamefont {Stamm}, \citenamefont {Kachel},
  \citenamefont {Pontius}, \citenamefont {D\"urr}, \citenamefont {Ostler},
  \citenamefont {Barker}, \citenamefont {Evans}, \citenamefont {Chantrell},
  \citenamefont {Tsukamoto}, \citenamefont {Itoh}, \citenamefont {Kirilyuk},
  \citenamefont {Rasing},\ and\ \citenamefont {Kimel}}]{radu_transient_2011}%
  \BibitemOpen
  \bibfield  {author} {\bibinfo {author} {\bibfnamefont {I.}~\bibnamefont
  {Radu}}, \bibinfo {author} {\bibfnamefont {K.}~\bibnamefont {Vahaplar}},
  \bibinfo {author} {\bibfnamefont {C.}~\bibnamefont {Stamm}}, \bibinfo
  {author} {\bibfnamefont {T.}~\bibnamefont {Kachel}}, \bibinfo {author}
  {\bibfnamefont {N.}~\bibnamefont {Pontius}}, \bibinfo {author} {\bibfnamefont
  {H.~A.}\ \bibnamefont {D\"urr}}, \bibinfo {author} {\bibfnamefont {T.~A.}\
  \bibnamefont {Ostler}}, \bibinfo {author} {\bibfnamefont {J.}~\bibnamefont
  {Barker}}, \bibinfo {author} {\bibfnamefont {R.~F.~L.}\ \bibnamefont
  {Evans}}, \bibinfo {author} {\bibfnamefont {R.~W.}\ \bibnamefont
  {Chantrell}}, \bibinfo {author} {\bibfnamefont {A.}~\bibnamefont
  {Tsukamoto}}, \bibinfo {author} {\bibfnamefont {A.}~\bibnamefont {Itoh}},
  \bibinfo {author} {\bibfnamefont {A.}~\bibnamefont {Kirilyuk}}, \bibinfo
  {author} {\bibfnamefont {T.}~\bibnamefont {Rasing}}, \ and\ \bibinfo {author}
  {\bibfnamefont {A.~V.}\ \bibnamefont {Kimel}},\ }\href {\doibase
  10.1038/nature09901} {\bibfield  {journal} {\bibinfo  {journal} {Nature}\
  }\textbf {\bibinfo {volume} {472}},\ \bibinfo {pages} {205} (\bibinfo {year}
  {2011})}\BibitemShut {NoStop}%
\bibitem [{\citenamefont {Vodungbo}\ \emph {et~al.}(2016)\citenamefont
  {Vodungbo}, \citenamefont {Tudu}, \citenamefont {Perron}, \citenamefont
  {Delaunay}, \citenamefont {M\"uller}, \citenamefont {Berntsen}, \citenamefont
  {Gr\"ubel}, \citenamefont {Malinowski}, \citenamefont {Weier}, \citenamefont
  {Gautier}, \citenamefont {Lambert}, \citenamefont {Zeitoun}, \citenamefont
  {Gutt}, \citenamefont {Jal}, \citenamefont {Reid}, \citenamefont {Granitzka},
  \citenamefont {Jaouen}, \citenamefont {Dakovski}, \citenamefont {Moeller},
  \citenamefont {Minitti}, \citenamefont {Mitra}, \citenamefont {Carron},
  \citenamefont {Pfau}, \citenamefont {von Korff~Schmising}, \citenamefont
  {Schneider}, \citenamefont {Eisebitt},\ and\ \citenamefont
  {L\"uning}}]{vodungbo_indirect_2016}%
  \BibitemOpen
  \bibfield  {author} {\bibinfo {author} {\bibfnamefont {B.}~\bibnamefont
  {Vodungbo}}, \bibinfo {author} {\bibfnamefont {B.}~\bibnamefont {Tudu}},
  \bibinfo {author} {\bibfnamefont {J.}~\bibnamefont {Perron}}, \bibinfo
  {author} {\bibfnamefont {R.}~\bibnamefont {Delaunay}}, \bibinfo {author}
  {\bibfnamefont {L.}~\bibnamefont {M\"uller}}, \bibinfo {author}
  {\bibfnamefont {M.~H.}\ \bibnamefont {Berntsen}}, \bibinfo {author}
  {\bibfnamefont {G.}~\bibnamefont {Gr\"ubel}}, \bibinfo {author}
  {\bibfnamefont {G.}~\bibnamefont {Malinowski}}, \bibinfo {author}
  {\bibfnamefont {C.}~\bibnamefont {Weier}}, \bibinfo {author} {\bibfnamefont
  {J.}~\bibnamefont {Gautier}}, \bibinfo {author} {\bibfnamefont
  {G.}~\bibnamefont {Lambert}}, \bibinfo {author} {\bibfnamefont
  {P.}~\bibnamefont {Zeitoun}}, \bibinfo {author} {\bibfnamefont
  {C.}~\bibnamefont {Gutt}}, \bibinfo {author} {\bibfnamefont {E.}~\bibnamefont
  {Jal}}, \bibinfo {author} {\bibfnamefont {A.~H.}\ \bibnamefont {Reid}},
  \bibinfo {author} {\bibfnamefont {P.~W.}\ \bibnamefont {Granitzka}}, \bibinfo
  {author} {\bibfnamefont {N.}~\bibnamefont {Jaouen}}, \bibinfo {author}
  {\bibfnamefont {G.~L.}\ \bibnamefont {Dakovski}}, \bibinfo {author}
  {\bibfnamefont {S.}~\bibnamefont {Moeller}}, \bibinfo {author} {\bibfnamefont
  {M.~P.}\ \bibnamefont {Minitti}}, \bibinfo {author} {\bibfnamefont
  {A.}~\bibnamefont {Mitra}}, \bibinfo {author} {\bibfnamefont
  {S.}~\bibnamefont {Carron}}, \bibinfo {author} {\bibfnamefont
  {B.}~\bibnamefont {Pfau}}, \bibinfo {author} {\bibfnamefont {C.}~\bibnamefont
  {von Korff~Schmising}}, \bibinfo {author} {\bibfnamefont {M.}~\bibnamefont
  {Schneider}}, \bibinfo {author} {\bibfnamefont {S.}~\bibnamefont {Eisebitt}},
  \ and\ \bibinfo {author} {\bibfnamefont {J.}~\bibnamefont {L\"uning}},\
  }\href {\doibase 10.1038/srep18970} {\bibfield  {journal} {\bibinfo
  {journal} {Scientific Reports}\ }\textbf {\bibinfo {volume} {6}},\ \bibinfo
  {pages} {18970} (\bibinfo {year} {2016})}\BibitemShut {NoStop}%
\bibitem [{\citenamefont {Gerber}\ \emph {et~al.}(2017)\citenamefont {Gerber},
  \citenamefont {Yang}, \citenamefont {Zhu}, \citenamefont {Soifer},
  \citenamefont {Sobota}, \citenamefont {Rebec}, \citenamefont {Lee},
  \citenamefont {Jia}, \citenamefont {Moritz}, \citenamefont {Jia},
  \citenamefont {Gauthier}, \citenamefont {Li}, \citenamefont {Leuenberger},
  \citenamefont {Zhang}, \citenamefont {Chaix}, \citenamefont {Li},
  \citenamefont {Jang}, \citenamefont {Lee}, \citenamefont {Yi}, \citenamefont
  {Dakovski}, \citenamefont {Song}, \citenamefont {Glownia}, \citenamefont
  {Nelson}, \citenamefont {Kim}, \citenamefont {Chuang}, \citenamefont
  {Hussain}, \citenamefont {Moore}, \citenamefont {Devereaux}, \citenamefont
  {Lee}, \citenamefont {Kirchmann},\ and\ \citenamefont
  {Shen}}]{gerber_femtosecond_2017}%
  \BibitemOpen
  \bibfield  {author} {\bibinfo {author} {\bibfnamefont {S.}~\bibnamefont
  {Gerber}}, \bibinfo {author} {\bibfnamefont {S.-L.}\ \bibnamefont {Yang}},
  \bibinfo {author} {\bibfnamefont {D.}~\bibnamefont {Zhu}}, \bibinfo {author}
  {\bibfnamefont {H.}~\bibnamefont {Soifer}}, \bibinfo {author} {\bibfnamefont
  {J.~A.}\ \bibnamefont {Sobota}}, \bibinfo {author} {\bibfnamefont
  {S.}~\bibnamefont {Rebec}}, \bibinfo {author} {\bibfnamefont {J.~J.}\
  \bibnamefont {Lee}}, \bibinfo {author} {\bibfnamefont {T.}~\bibnamefont
  {Jia}}, \bibinfo {author} {\bibfnamefont {B.}~\bibnamefont {Moritz}},
  \bibinfo {author} {\bibfnamefont {C.}~\bibnamefont {Jia}}, \bibinfo {author}
  {\bibfnamefont {A.}~\bibnamefont {Gauthier}}, \bibinfo {author}
  {\bibfnamefont {Y.}~\bibnamefont {Li}}, \bibinfo {author} {\bibfnamefont
  {D.}~\bibnamefont {Leuenberger}}, \bibinfo {author} {\bibfnamefont
  {Y.}~\bibnamefont {Zhang}}, \bibinfo {author} {\bibfnamefont
  {L.}~\bibnamefont {Chaix}}, \bibinfo {author} {\bibfnamefont
  {W.}~\bibnamefont {Li}}, \bibinfo {author} {\bibfnamefont {H.}~\bibnamefont
  {Jang}}, \bibinfo {author} {\bibfnamefont {J.-S.}\ \bibnamefont {Lee}},
  \bibinfo {author} {\bibfnamefont {M.}~\bibnamefont {Yi}}, \bibinfo {author}
  {\bibfnamefont {G.~L.}\ \bibnamefont {Dakovski}}, \bibinfo {author}
  {\bibfnamefont {S.}~\bibnamefont {Song}}, \bibinfo {author} {\bibfnamefont
  {J.~M.}\ \bibnamefont {Glownia}}, \bibinfo {author} {\bibfnamefont
  {S.}~\bibnamefont {Nelson}}, \bibinfo {author} {\bibfnamefont {K.~W.}\
  \bibnamefont {Kim}}, \bibinfo {author} {\bibfnamefont {Y.-D.}\ \bibnamefont
  {Chuang}}, \bibinfo {author} {\bibfnamefont {Z.}~\bibnamefont {Hussain}},
  \bibinfo {author} {\bibfnamefont {R.~G.}\ \bibnamefont {Moore}}, \bibinfo
  {author} {\bibfnamefont {T.~P.}\ \bibnamefont {Devereaux}}, \bibinfo {author}
  {\bibfnamefont {W.-S.}\ \bibnamefont {Lee}}, \bibinfo {author} {\bibfnamefont
  {P.~S.}\ \bibnamefont {Kirchmann}}, \ and\ \bibinfo {author} {\bibfnamefont
  {Z.-X.}\ \bibnamefont {Shen}},\ }\href {\doibase 10.1126/science.aak9946}
  {\bibfield  {journal} {\bibinfo  {journal} {Science}\ }\textbf {\bibinfo
  {volume} {357}},\ \bibinfo {pages} {71} (\bibinfo {year} {2017})}\BibitemShut
  {NoStop}%
\bibitem [{\citenamefont {Chaix}\ \emph {et~al.}(2017)\citenamefont {Chaix},
  \citenamefont {Ghiringhelli}, \citenamefont {Peng}, \citenamefont
  {Hashimoto}, \citenamefont {Moritz}, \citenamefont {Kummer}, \citenamefont
  {Brookes}, \citenamefont {He}, \citenamefont {Chen}, \citenamefont {Ishida},
  \citenamefont {Yoshida}, \citenamefont {Eisaki}, \citenamefont {Salluzzo},
  \citenamefont {Braicovich}, \citenamefont {Shen}, \citenamefont {Devereaux},\
  and\ \citenamefont {Lee}}]{chaix_dispersive_2017}%
  \BibitemOpen
  \bibfield  {author} {\bibinfo {author} {\bibfnamefont {L.}~\bibnamefont
  {Chaix}}, \bibinfo {author} {\bibfnamefont {G.}~\bibnamefont {Ghiringhelli}},
  \bibinfo {author} {\bibfnamefont {Y.~Y.}\ \bibnamefont {Peng}}, \bibinfo
  {author} {\bibfnamefont {M.}~\bibnamefont {Hashimoto}}, \bibinfo {author}
  {\bibfnamefont {B.}~\bibnamefont {Moritz}}, \bibinfo {author} {\bibfnamefont
  {K.}~\bibnamefont {Kummer}}, \bibinfo {author} {\bibfnamefont {N.~B.}\
  \bibnamefont {Brookes}}, \bibinfo {author} {\bibfnamefont {Y.}~\bibnamefont
  {He}}, \bibinfo {author} {\bibfnamefont {S.}~\bibnamefont {Chen}}, \bibinfo
  {author} {\bibfnamefont {S.}~\bibnamefont {Ishida}}, \bibinfo {author}
  {\bibfnamefont {Y.}~\bibnamefont {Yoshida}}, \bibinfo {author} {\bibfnamefont
  {H.}~\bibnamefont {Eisaki}}, \bibinfo {author} {\bibfnamefont
  {M.}~\bibnamefont {Salluzzo}}, \bibinfo {author} {\bibfnamefont
  {L.}~\bibnamefont {Braicovich}}, \bibinfo {author} {\bibfnamefont {Z.-X.}\
  \bibnamefont {Shen}}, \bibinfo {author} {\bibfnamefont {T.~P.}\ \bibnamefont
  {Devereaux}}, \ and\ \bibinfo {author} {\bibfnamefont {W.-S.}\ \bibnamefont
  {Lee}},\ }\href {http://dx.doi.org/10.1038/nphys4157} {\bibfield  {journal}
  {\bibinfo  {journal} {Nature Physics}\ }\textbf {\bibinfo {volume} {13}},\
  \bibinfo {pages} {952} (\bibinfo {year} {2017})}\BibitemShut {NoStop}%
\bibitem [{\citenamefont {Wang}\ \emph {et~al.}(2012)\citenamefont {Wang},
  \citenamefont {Zhu}, \citenamefont {Wu}, \citenamefont {Graves},
  \citenamefont {Schaffert}, \citenamefont {Rander}, \citenamefont {M\"uller},
  \citenamefont {Vodungbo}, \citenamefont {Baumier}, \citenamefont {Bernstein},
  \citenamefont {Br\"auer}, \citenamefont {Cros}, \citenamefont {de~Jong},
  \citenamefont {Delaunay}, \citenamefont {Fognini}, \citenamefont {Kukreja},
  \citenamefont {Lee}, \citenamefont {L\'opez-Flores}, \citenamefont {Mohanty},
  \citenamefont {Pfau}, \citenamefont {Popescu}, \citenamefont {Sacchi},
  \citenamefont {Sardinha}, \citenamefont {Sirotti}, \citenamefont {Zeitoun},
  \citenamefont {Messerschmidt}, \citenamefont {Turner}, \citenamefont
  {Schlotter}, \citenamefont {Hellwig}, \citenamefont {Mattana}, \citenamefont
  {Jaouen}, \citenamefont {Fortuna}, \citenamefont {Acremann}, \citenamefont
  {Gutt}, \citenamefont {D\"urr}, \citenamefont {Beaurepaire}, \citenamefont
  {Boeglin}, \citenamefont {Eisebitt}, \citenamefont {Gr\"ubel}, \citenamefont
  {L\"uning}, \citenamefont {St\"ohr},\ and\ \citenamefont
  {Scherz}}]{wang_femtosecond_2012}%
  \BibitemOpen
  \bibfield  {author} {\bibinfo {author} {\bibfnamefont {T.}~\bibnamefont
  {Wang}}, \bibinfo {author} {\bibfnamefont {D.}~\bibnamefont {Zhu}}, \bibinfo
  {author} {\bibfnamefont {B.}~\bibnamefont {Wu}}, \bibinfo {author}
  {\bibfnamefont {C.}~\bibnamefont {Graves}}, \bibinfo {author} {\bibfnamefont
  {S.}~\bibnamefont {Schaffert}}, \bibinfo {author} {\bibfnamefont
  {T.}~\bibnamefont {Rander}}, \bibinfo {author} {\bibfnamefont
  {L.}~\bibnamefont {M\"uller}}, \bibinfo {author} {\bibfnamefont
  {B.}~\bibnamefont {Vodungbo}}, \bibinfo {author} {\bibfnamefont
  {C.}~\bibnamefont {Baumier}}, \bibinfo {author} {\bibfnamefont
  {D.}~\bibnamefont {Bernstein}}, \bibinfo {author} {\bibfnamefont
  {B.}~\bibnamefont {Br\"auer}}, \bibinfo {author} {\bibfnamefont
  {V.}~\bibnamefont {Cros}}, \bibinfo {author} {\bibfnamefont {S.}~\bibnamefont
  {de~Jong}}, \bibinfo {author} {\bibfnamefont {R.}~\bibnamefont {Delaunay}},
  \bibinfo {author} {\bibfnamefont {A.}~\bibnamefont {Fognini}}, \bibinfo
  {author} {\bibfnamefont {R.}~\bibnamefont {Kukreja}}, \bibinfo {author}
  {\bibfnamefont {S.}~\bibnamefont {Lee}}, \bibinfo {author} {\bibfnamefont
  {V.}~\bibnamefont {L\'opez-Flores}}, \bibinfo {author} {\bibfnamefont
  {J.}~\bibnamefont {Mohanty}}, \bibinfo {author} {\bibfnamefont
  {B.}~\bibnamefont {Pfau}}, \bibinfo {author} {\bibfnamefont {H.}~\bibnamefont
  {Popescu}}, \bibinfo {author} {\bibfnamefont {M.}~\bibnamefont {Sacchi}},
  \bibinfo {author} {\bibfnamefont {A.}~\bibnamefont {Sardinha}}, \bibinfo
  {author} {\bibfnamefont {F.}~\bibnamefont {Sirotti}}, \bibinfo {author}
  {\bibfnamefont {P.}~\bibnamefont {Zeitoun}}, \bibinfo {author} {\bibfnamefont
  {M.}~\bibnamefont {Messerschmidt}}, \bibinfo {author} {\bibfnamefont
  {J.}~\bibnamefont {Turner}}, \bibinfo {author} {\bibfnamefont
  {W.}~\bibnamefont {Schlotter}}, \bibinfo {author} {\bibfnamefont
  {O.}~\bibnamefont {Hellwig}}, \bibinfo {author} {\bibfnamefont
  {R.}~\bibnamefont {Mattana}}, \bibinfo {author} {\bibfnamefont
  {N.}~\bibnamefont {Jaouen}}, \bibinfo {author} {\bibfnamefont
  {F.}~\bibnamefont {Fortuna}}, \bibinfo {author} {\bibfnamefont
  {Y.}~\bibnamefont {Acremann}}, \bibinfo {author} {\bibfnamefont
  {C.}~\bibnamefont {Gutt}}, \bibinfo {author} {\bibfnamefont {H.}~\bibnamefont
  {D\"urr}}, \bibinfo {author} {\bibfnamefont {E.}~\bibnamefont {Beaurepaire}},
  \bibinfo {author} {\bibfnamefont {C.}~\bibnamefont {Boeglin}}, \bibinfo
  {author} {\bibfnamefont {S.}~\bibnamefont {Eisebitt}}, \bibinfo {author}
  {\bibfnamefont {G.}~\bibnamefont {Gr\"ubel}}, \bibinfo {author}
  {\bibfnamefont {J.}~\bibnamefont {L\"uning}}, \bibinfo {author}
  {\bibfnamefont {J.}~\bibnamefont {St\"ohr}}, \ and\ \bibinfo {author}
  {\bibfnamefont {A.}~\bibnamefont {Scherz}},\ }\href {\doibase
  10.1103/PhysRevLett.108.267403} {\bibfield  {journal} {\bibinfo  {journal}
  {Physical Review Letters}\ }\textbf {\bibinfo {volume} {108}},\ \bibinfo
  {pages} {267403} (\bibinfo {year} {2012})}\BibitemShut {NoStop}%
\bibitem [{\citenamefont {Buzzi}\ \emph {et~al.}(2017)\citenamefont {Buzzi},
  \citenamefont {Makita}, \citenamefont {Howald}, \citenamefont {Kleibert},
  \citenamefont {Vodungbo}, \citenamefont {Maldonado}, \citenamefont {Raabe},
  \citenamefont {Jaouen}, \citenamefont {Redlin}, \citenamefont {Tiedtke},
  \citenamefont {Oppeneer}, \citenamefont {David}, \citenamefont {Nolting},\
  and\ \citenamefont {L\"uning}}]{buzzi_single-shot_2017}%
  \BibitemOpen
  \bibfield  {author} {\bibinfo {author} {\bibfnamefont {M.}~\bibnamefont
  {Buzzi}}, \bibinfo {author} {\bibfnamefont {M.}~\bibnamefont {Makita}},
  \bibinfo {author} {\bibfnamefont {L.}~\bibnamefont {Howald}}, \bibinfo
  {author} {\bibfnamefont {A.}~\bibnamefont {Kleibert}}, \bibinfo {author}
  {\bibfnamefont {B.}~\bibnamefont {Vodungbo}}, \bibinfo {author}
  {\bibfnamefont {P.}~\bibnamefont {Maldonado}}, \bibinfo {author}
  {\bibfnamefont {J.}~\bibnamefont {Raabe}}, \bibinfo {author} {\bibfnamefont
  {N.}~\bibnamefont {Jaouen}}, \bibinfo {author} {\bibfnamefont
  {H.}~\bibnamefont {Redlin}}, \bibinfo {author} {\bibfnamefont
  {K.}~\bibnamefont {Tiedtke}}, \bibinfo {author} {\bibfnamefont {P.~M.}\
  \bibnamefont {Oppeneer}}, \bibinfo {author} {\bibfnamefont {C.}~\bibnamefont
  {David}}, \bibinfo {author} {\bibfnamefont {F.}~\bibnamefont {Nolting}}, \
  and\ \bibinfo {author} {\bibfnamefont {J.}~\bibnamefont {L\"uning}},\ }\href
  {\doibase 10.1038/s41598-017-07069-z} {\bibfield  {journal} {\bibinfo
  {journal} {Scientific Reports}\ }\textbf {\bibinfo {volume} {7}},\ \bibinfo
  {pages} {7253} (\bibinfo {year} {2017})}\BibitemShut {NoStop}%
\bibitem [{\citenamefont {Chen}\ \emph {et~al.}(1995)\citenamefont {Chen},
  \citenamefont {Idzerda}, \citenamefont {Lin}, \citenamefont {Smith},
  \citenamefont {Meigs}, \citenamefont {Chaban}, \citenamefont {Ho},
  \citenamefont {Pellegrin},\ and\ \citenamefont
  {Sette}}]{chen_experimental_1995}%
  \BibitemOpen
  \bibfield  {author} {\bibinfo {author} {\bibfnamefont {C.~T.}\ \bibnamefont
  {Chen}}, \bibinfo {author} {\bibfnamefont {Y.~U.}\ \bibnamefont {Idzerda}},
  \bibinfo {author} {\bibfnamefont {H.-J.}\ \bibnamefont {Lin}}, \bibinfo
  {author} {\bibfnamefont {N.~V.}\ \bibnamefont {Smith}}, \bibinfo {author}
  {\bibfnamefont {G.}~\bibnamefont {Meigs}}, \bibinfo {author} {\bibfnamefont
  {E.}~\bibnamefont {Chaban}}, \bibinfo {author} {\bibfnamefont {G.~H.}\
  \bibnamefont {Ho}}, \bibinfo {author} {\bibfnamefont {E.}~\bibnamefont
  {Pellegrin}}, \ and\ \bibinfo {author} {\bibfnamefont {F.}~\bibnamefont
  {Sette}},\ }\href {\doibase 10.1103/PhysRevLett.75.152} {\bibfield  {journal}
  {\bibinfo  {journal} {Physical Review Letters}\ }\textbf {\bibinfo {volume}
  {75}},\ \bibinfo {pages} {152} (\bibinfo {year} {1995})}\BibitemShut
  {NoStop}%
\bibitem [{\citenamefont {Valencia}\ \emph {et~al.}(2006)\citenamefont
  {Valencia}, \citenamefont {Gaupp}, \citenamefont {Gudat}, \citenamefont
  {Mertins}, \citenamefont {Oppeneer}, \citenamefont {Abramsohn},\ and\
  \citenamefont {Schneider}}]{valencia_faraday_2006}%
  \BibitemOpen
  \bibfield  {author} {\bibinfo {author} {\bibfnamefont {S.}~\bibnamefont
  {Valencia}}, \bibinfo {author} {\bibfnamefont {A.}~\bibnamefont {Gaupp}},
  \bibinfo {author} {\bibfnamefont {W.}~\bibnamefont {Gudat}}, \bibinfo
  {author} {\bibfnamefont {H.-C.}\ \bibnamefont {Mertins}}, \bibinfo {author}
  {\bibfnamefont {P.~M.}\ \bibnamefont {Oppeneer}}, \bibinfo {author}
  {\bibfnamefont {D.}~\bibnamefont {Abramsohn}}, \ and\ \bibinfo {author}
  {\bibfnamefont {C.~M.}\ \bibnamefont {Schneider}},\ }\href {\doibase
  10.1088/1367-2630/8/10/254} {\bibfield  {journal} {\bibinfo  {journal} {New
  Journal of Physics}\ }\textbf {\bibinfo {volume} {8}},\ \bibinfo {pages}
  {254} (\bibinfo {year} {2006})}\BibitemShut {NoStop}%
\bibitem [{\citenamefont {Ramasesha}\ \emph {et~al.}(2016)\citenamefont
  {Ramasesha}, \citenamefont {Leone},\ and\ \citenamefont
  {Neumark}}]{ramasesha_real-time_2016}%
  \BibitemOpen
  \bibfield  {author} {\bibinfo {author} {\bibfnamefont {K.}~\bibnamefont
  {Ramasesha}}, \bibinfo {author} {\bibfnamefont {S.~R.}\ \bibnamefont
  {Leone}}, \ and\ \bibinfo {author} {\bibfnamefont {D.~M.}\ \bibnamefont
  {Neumark}},\ }\href {\doibase 10.1146/annurev-physchem-040215-112025}
  {\bibfield  {journal} {\bibinfo  {journal} {Annual Review of Physical
  Chemistry}\ }\textbf {\bibinfo {volume} {67}},\ \bibinfo {pages} {41}
  (\bibinfo {year} {2016})}\BibitemShut {NoStop}%
\bibitem [{\citenamefont {Nguyen}\ \emph {et~al.}(2016)\citenamefont {Nguyen},
  \citenamefont {Zhang}, \citenamefont {Manthiram}, \citenamefont {Ye},
  \citenamefont {Lomont}, \citenamefont {Harris}, \citenamefont {Weller},\ and\
  \citenamefont {Alivisatos}}]{nguyen_study_2016}%
  \BibitemOpen
  \bibfield  {author} {\bibinfo {author} {\bibfnamefont {S.~C.}\ \bibnamefont
  {Nguyen}}, \bibinfo {author} {\bibfnamefont {Q.}~\bibnamefont {Zhang}},
  \bibinfo {author} {\bibfnamefont {K.}~\bibnamefont {Manthiram}}, \bibinfo
  {author} {\bibfnamefont {X.}~\bibnamefont {Ye}}, \bibinfo {author}
  {\bibfnamefont {J.~P.}\ \bibnamefont {Lomont}}, \bibinfo {author}
  {\bibfnamefont {C.~B.}\ \bibnamefont {Harris}}, \bibinfo {author}
  {\bibfnamefont {H.}~\bibnamefont {Weller}}, \ and\ \bibinfo {author}
  {\bibfnamefont {A.~P.}\ \bibnamefont {Alivisatos}},\ }\href {\doibase
  10.1021/acsnano.5b06623} {\bibfield  {journal} {\bibinfo  {journal} {ACS
  Nano}\ }\textbf {\bibinfo {volume} {10}},\ \bibinfo {pages} {2144} (\bibinfo
  {year} {2016})}\BibitemShut {NoStop}%
\bibitem [{\citenamefont {Vaida}\ \emph {et~al.}(2018)\citenamefont {Vaida},
  \citenamefont {Marsh},\ and\ \citenamefont {Leone}}]{vaida_nonmetal_2018}%
  \BibitemOpen
  \bibfield  {author} {\bibinfo {author} {\bibfnamefont {M.~E.}\ \bibnamefont
  {Vaida}}, \bibinfo {author} {\bibfnamefont {B.~M.}\ \bibnamefont {Marsh}}, \
  and\ \bibinfo {author} {\bibfnamefont {S.~R.}\ \bibnamefont {Leone}},\ }\href
  {\doibase 10.1021/acs.nanolett.8b00700} {\bibfield  {journal} {\bibinfo
  {journal} {Nano Letters}\ }\textbf {\bibinfo {volume} {18}},\ \bibinfo
  {pages} {4107} (\bibinfo {year} {2018})}\BibitemShut {NoStop}%
\bibitem [{\citenamefont {Beaurepaire}\ \emph {et~al.}(1996)\citenamefont
  {Beaurepaire}, \citenamefont {Merle}, \citenamefont {Daunois},\ and\
  \citenamefont {Bigot}}]{beaurepaire_ultrafast_1996}%
  \BibitemOpen
  \bibfield  {author} {\bibinfo {author} {\bibfnamefont {E.}~\bibnamefont
  {Beaurepaire}}, \bibinfo {author} {\bibfnamefont {J.-C.}\ \bibnamefont
  {Merle}}, \bibinfo {author} {\bibfnamefont {A.}~\bibnamefont {Daunois}}, \
  and\ \bibinfo {author} {\bibfnamefont {J.-Y.}\ \bibnamefont {Bigot}},\ }\href
  {\doibase 10.1103/PhysRevLett.76.4250} {\bibfield  {journal} {\bibinfo
  {journal} {Phys. Rev. Lett.}\ }\textbf {\bibinfo {volume} {76}},\ \bibinfo
  {pages} {4250} (\bibinfo {year} {1996})}\BibitemShut {NoStop}%
\bibitem [{\citenamefont {Stanciu}\ \emph {et~al.}(2007)\citenamefont
  {Stanciu}, \citenamefont {Hansteen}, \citenamefont {Kimel}, \citenamefont
  {Kirilyuk}, \citenamefont {Tsukamoto}, \citenamefont {Itoh},\ and\
  \citenamefont {Rasing}}]{stanciu_all-optical_2007}%
  \BibitemOpen
  \bibfield  {author} {\bibinfo {author} {\bibfnamefont {C.~D.}\ \bibnamefont
  {Stanciu}}, \bibinfo {author} {\bibfnamefont {F.}~\bibnamefont {Hansteen}},
  \bibinfo {author} {\bibfnamefont {A.~V.}\ \bibnamefont {Kimel}}, \bibinfo
  {author} {\bibfnamefont {A.}~\bibnamefont {Kirilyuk}}, \bibinfo {author}
  {\bibfnamefont {A.}~\bibnamefont {Tsukamoto}}, \bibinfo {author}
  {\bibfnamefont {A.}~\bibnamefont {Itoh}}, \ and\ \bibinfo {author}
  {\bibfnamefont {T.}~\bibnamefont {Rasing}},\ }\href {\doibase
  10.1103/PhysRevLett.99.047601} {\bibfield  {journal} {\bibinfo  {journal}
  {Physical Review Letters}\ }\textbf {\bibinfo {volume} {99}},\ \bibinfo
  {pages} {047601} (\bibinfo {year} {2007})}\BibitemShut {NoStop}%
\bibitem [{\citenamefont {Capotondi}\ \emph {et~al.}(2013)\citenamefont
  {Capotondi}, \citenamefont {Pedersoli}, \citenamefont {Mahne}, \citenamefont
  {Menk}, \citenamefont {Passos}, \citenamefont {Raimondi}, \citenamefont
  {Svetina}, \citenamefont {Sandrin}, \citenamefont {Zangrando}, \citenamefont
  {Kiskinova}, \citenamefont {Bajt}, \citenamefont {Barthelmess}, \citenamefont
  {Fleckenstein}, \citenamefont {Chapman}, \citenamefont {Schulz},
  \citenamefont {Bach}, \citenamefont {Fr\"omter}, \citenamefont {Schleitzer},
  \citenamefont {M\"uller}, \citenamefont {Gutt},\ and\ \citenamefont
  {Gr\"ubel}}]{capotondi_invited_2013}%
  \BibitemOpen
  \bibfield  {author} {\bibinfo {author} {\bibfnamefont {F.}~\bibnamefont
  {Capotondi}}, \bibinfo {author} {\bibfnamefont {E.}~\bibnamefont
  {Pedersoli}}, \bibinfo {author} {\bibfnamefont {N.}~\bibnamefont {Mahne}},
  \bibinfo {author} {\bibfnamefont {R.~H.}\ \bibnamefont {Menk}}, \bibinfo
  {author} {\bibfnamefont {G.}~\bibnamefont {Passos}}, \bibinfo {author}
  {\bibfnamefont {L.}~\bibnamefont {Raimondi}}, \bibinfo {author}
  {\bibfnamefont {C.}~\bibnamefont {Svetina}}, \bibinfo {author} {\bibfnamefont
  {G.}~\bibnamefont {Sandrin}}, \bibinfo {author} {\bibfnamefont
  {M.}~\bibnamefont {Zangrando}}, \bibinfo {author} {\bibfnamefont
  {M.}~\bibnamefont {Kiskinova}}, \bibinfo {author} {\bibfnamefont
  {S.}~\bibnamefont {Bajt}}, \bibinfo {author} {\bibfnamefont {M.}~\bibnamefont
  {Barthelmess}}, \bibinfo {author} {\bibfnamefont {H.}~\bibnamefont
  {Fleckenstein}}, \bibinfo {author} {\bibfnamefont {H.~N.}\ \bibnamefont
  {Chapman}}, \bibinfo {author} {\bibfnamefont {J.}~\bibnamefont {Schulz}},
  \bibinfo {author} {\bibfnamefont {J.}~\bibnamefont {Bach}}, \bibinfo {author}
  {\bibfnamefont {R.}~\bibnamefont {Fr\"omter}}, \bibinfo {author}
  {\bibfnamefont {S.}~\bibnamefont {Schleitzer}}, \bibinfo {author}
  {\bibfnamefont {L.}~\bibnamefont {M\"uller}}, \bibinfo {author}
  {\bibfnamefont {C.}~\bibnamefont {Gutt}}, \ and\ \bibinfo {author}
  {\bibfnamefont {G.}~\bibnamefont {Gr\"ubel}},\ }\href {\doibase
  10.1063/1.4807157} {\bibfield  {journal} {\bibinfo  {journal} {Review of
  Scientific Instruments}\ }\textbf {\bibinfo {volume} {84}},\ \bibinfo {pages}
  {051301} (\bibinfo {year} {2013})}\BibitemShut {NoStop}%
\bibitem [{\citenamefont {Allaria}\ \emph {et~al.}(2012)\citenamefont
  {Allaria}, \citenamefont {Appio}, \citenamefont {Badano}, \citenamefont
  {Barletta}, \citenamefont {Bassanese}, \citenamefont {Biedron}, \citenamefont
  {Borga}, \citenamefont {Busetto}, \citenamefont {Castronovo}, \citenamefont
  {Cinquegrana}, \citenamefont {Cleva}, \citenamefont {Cocco}, \citenamefont
  {Cornacchia}, \citenamefont {Craievich}, \citenamefont {Cudin}, \citenamefont
  {D'Auria}, \citenamefont {Dal~Forno}, \citenamefont {Danailov}, \citenamefont
  {De~Monte}, \citenamefont {De~Ninno}, \citenamefont {Delgiusto},
  \citenamefont {Demidovich}, \citenamefont {Di~Mitri}, \citenamefont
  {Diviacco}, \citenamefont {Fabris}, \citenamefont {Fabris}, \citenamefont
  {Fawley}, \citenamefont {Ferianis}, \citenamefont {Ferrari}, \citenamefont
  {Ferry}, \citenamefont {Froehlich}, \citenamefont {Furlan}, \citenamefont
  {Gaio}, \citenamefont {Gelmetti}, \citenamefont {Giannessi}, \citenamefont
  {Giannini}, \citenamefont {Gobessi}, \citenamefont {Ivanov}, \citenamefont
  {Karantzoulis}, \citenamefont {Lonza}, \citenamefont {Lutman}, \citenamefont
  {Mahieu}, \citenamefont {Milloch}, \citenamefont {Milton}, \citenamefont
  {Musardo}, \citenamefont {Nikolov}, \citenamefont {Noe}, \citenamefont
  {Parmigiani}, \citenamefont {Penco}, \citenamefont {Petronio}, \citenamefont
  {Pivetta}, \citenamefont {Predonzani}, \citenamefont {Rossi}, \citenamefont
  {Rumiz}, \citenamefont {Salom}, \citenamefont {Scafuri}, \citenamefont
  {Serpico}, \citenamefont {Sigalotti}, \citenamefont {Spampinati},
  \citenamefont {Spezzani}, \citenamefont {Svandrlik}, \citenamefont {Svetina},
  \citenamefont {Tazzari}, \citenamefont {Trovo}, \citenamefont {Umer},
  \citenamefont {Vascotto}, \citenamefont {Veronese}, \citenamefont
  {Visintini}, \citenamefont {Zaccaria}, \citenamefont {Zangrando},\ and\
  \citenamefont {Zangrando}}]{allaria_highly_2012}%
  \BibitemOpen
  \bibfield  {author} {\bibinfo {author} {\bibfnamefont {E.}~\bibnamefont
  {Allaria}}, \bibinfo {author} {\bibfnamefont {R.}~\bibnamefont {Appio}},
  \bibinfo {author} {\bibfnamefont {L.}~\bibnamefont {Badano}}, \bibinfo
  {author} {\bibfnamefont {W.~A.}\ \bibnamefont {Barletta}}, \bibinfo {author}
  {\bibfnamefont {S.}~\bibnamefont {Bassanese}}, \bibinfo {author}
  {\bibfnamefont {S.~G.}\ \bibnamefont {Biedron}}, \bibinfo {author}
  {\bibfnamefont {A.}~\bibnamefont {Borga}}, \bibinfo {author} {\bibfnamefont
  {E.}~\bibnamefont {Busetto}}, \bibinfo {author} {\bibfnamefont
  {D.}~\bibnamefont {Castronovo}}, \bibinfo {author} {\bibfnamefont
  {P.}~\bibnamefont {Cinquegrana}}, \bibinfo {author} {\bibfnamefont
  {S.}~\bibnamefont {Cleva}}, \bibinfo {author} {\bibfnamefont
  {D.}~\bibnamefont {Cocco}}, \bibinfo {author} {\bibfnamefont
  {M.}~\bibnamefont {Cornacchia}}, \bibinfo {author} {\bibfnamefont
  {P.}~\bibnamefont {Craievich}}, \bibinfo {author} {\bibfnamefont
  {I.}~\bibnamefont {Cudin}}, \bibinfo {author} {\bibfnamefont
  {G.}~\bibnamefont {D'Auria}}, \bibinfo {author} {\bibfnamefont
  {M.}~\bibnamefont {Dal~Forno}}, \bibinfo {author} {\bibfnamefont {M.~B.}\
  \bibnamefont {Danailov}}, \bibinfo {author} {\bibfnamefont {R.}~\bibnamefont
  {De~Monte}}, \bibinfo {author} {\bibfnamefont {G.}~\bibnamefont {De~Ninno}},
  \bibinfo {author} {\bibfnamefont {P.}~\bibnamefont {Delgiusto}}, \bibinfo
  {author} {\bibfnamefont {A.}~\bibnamefont {Demidovich}}, \bibinfo {author}
  {\bibfnamefont {S.}~\bibnamefont {Di~Mitri}}, \bibinfo {author}
  {\bibfnamefont {B.}~\bibnamefont {Diviacco}}, \bibinfo {author}
  {\bibfnamefont {A.}~\bibnamefont {Fabris}}, \bibinfo {author} {\bibfnamefont
  {R.}~\bibnamefont {Fabris}}, \bibinfo {author} {\bibfnamefont
  {W.}~\bibnamefont {Fawley}}, \bibinfo {author} {\bibfnamefont
  {M.}~\bibnamefont {Ferianis}}, \bibinfo {author} {\bibfnamefont
  {E.}~\bibnamefont {Ferrari}}, \bibinfo {author} {\bibfnamefont
  {S.}~\bibnamefont {Ferry}}, \bibinfo {author} {\bibfnamefont
  {L.}~\bibnamefont {Froehlich}}, \bibinfo {author} {\bibfnamefont
  {P.}~\bibnamefont {Furlan}}, \bibinfo {author} {\bibfnamefont
  {G.}~\bibnamefont {Gaio}}, \bibinfo {author} {\bibfnamefont {F.}~\bibnamefont
  {Gelmetti}}, \bibinfo {author} {\bibfnamefont {L.}~\bibnamefont {Giannessi}},
  \bibinfo {author} {\bibfnamefont {M.}~\bibnamefont {Giannini}}, \bibinfo
  {author} {\bibfnamefont {R.}~\bibnamefont {Gobessi}}, \bibinfo {author}
  {\bibfnamefont {R.}~\bibnamefont {Ivanov}}, \bibinfo {author} {\bibfnamefont
  {E.}~\bibnamefont {Karantzoulis}}, \bibinfo {author} {\bibfnamefont
  {M.}~\bibnamefont {Lonza}}, \bibinfo {author} {\bibfnamefont
  {A.}~\bibnamefont {Lutman}}, \bibinfo {author} {\bibfnamefont
  {B.}~\bibnamefont {Mahieu}}, \bibinfo {author} {\bibfnamefont
  {M.}~\bibnamefont {Milloch}}, \bibinfo {author} {\bibfnamefont {S.~V.}\
  \bibnamefont {Milton}}, \bibinfo {author} {\bibfnamefont {M.}~\bibnamefont
  {Musardo}}, \bibinfo {author} {\bibfnamefont {I.}~\bibnamefont {Nikolov}},
  \bibinfo {author} {\bibfnamefont {S.}~\bibnamefont {Noe}}, \bibinfo {author}
  {\bibfnamefont {F.}~\bibnamefont {Parmigiani}}, \bibinfo {author}
  {\bibfnamefont {G.}~\bibnamefont {Penco}}, \bibinfo {author} {\bibfnamefont
  {M.}~\bibnamefont {Petronio}}, \bibinfo {author} {\bibfnamefont
  {L.}~\bibnamefont {Pivetta}}, \bibinfo {author} {\bibfnamefont
  {M.}~\bibnamefont {Predonzani}}, \bibinfo {author} {\bibfnamefont
  {F.}~\bibnamefont {Rossi}}, \bibinfo {author} {\bibfnamefont
  {L.}~\bibnamefont {Rumiz}}, \bibinfo {author} {\bibfnamefont
  {A.}~\bibnamefont {Salom}}, \bibinfo {author} {\bibfnamefont
  {C.}~\bibnamefont {Scafuri}}, \bibinfo {author} {\bibfnamefont
  {C.}~\bibnamefont {Serpico}}, \bibinfo {author} {\bibfnamefont
  {P.}~\bibnamefont {Sigalotti}}, \bibinfo {author} {\bibfnamefont
  {S.}~\bibnamefont {Spampinati}}, \bibinfo {author} {\bibfnamefont
  {C.}~\bibnamefont {Spezzani}}, \bibinfo {author} {\bibfnamefont
  {M.}~\bibnamefont {Svandrlik}}, \bibinfo {author} {\bibfnamefont
  {C.}~\bibnamefont {Svetina}}, \bibinfo {author} {\bibfnamefont
  {S.}~\bibnamefont {Tazzari}}, \bibinfo {author} {\bibfnamefont
  {M.}~\bibnamefont {Trovo}}, \bibinfo {author} {\bibfnamefont
  {R.}~\bibnamefont {Umer}}, \bibinfo {author} {\bibfnamefont {A.}~\bibnamefont
  {Vascotto}}, \bibinfo {author} {\bibfnamefont {M.}~\bibnamefont {Veronese}},
  \bibinfo {author} {\bibfnamefont {R.}~\bibnamefont {Visintini}}, \bibinfo
  {author} {\bibfnamefont {M.}~\bibnamefont {Zaccaria}}, \bibinfo {author}
  {\bibfnamefont {D.}~\bibnamefont {Zangrando}}, \ and\ \bibinfo {author}
  {\bibfnamefont {M.}~\bibnamefont {Zangrando}},\ }\href {\doibase
  10.1038/nphoton.2012.233} {\bibfield  {journal} {\bibinfo  {journal} {Nature
  Photonics}\ }\textbf {\bibinfo {volume} {6}},\ \bibinfo {pages} {699}
  (\bibinfo {year} {2012})}\BibitemShut {NoStop}%
\bibitem [{\citenamefont {Theoretically}()}]{note}%
  \BibitemOpen
  \bibfield  {author} {\bibinfo {author} {\bibnamefont {Theoretically}},\
  }\href@noop {} {}\bibinfo {note} {The Co absorption edge is at 20.8 nm and we
  start the experiment at this wavelength but by measuring the magnetic
  contrast we observed that the best magnetic contrast was achieved for a
  wavelength of 20.5 nm}\BibitemShut {NoStop}%
\bibitem [{\citenamefont {Finetti}\ \emph {et~al.}(2017)\citenamefont
  {Finetti}, \citenamefont {H\"oppner}, \citenamefont {Allaria}, \citenamefont
  {Callegari}, \citenamefont {Capotondi}, \citenamefont {Cinquegrana},
  \citenamefont {Coreno}, \citenamefont {Cucini}, \citenamefont {Danailov},
  \citenamefont {Demidovich}, \citenamefont {De~Ninno}, \citenamefont
  {Di~Fraia}, \citenamefont {Feifel}, \citenamefont {Ferrari}, \citenamefont
  {Fr\"ohlich}, \citenamefont {Gauthier}, \citenamefont {Golz}, \citenamefont
  {Grazioli}, \citenamefont {Kai}, \citenamefont {Kurdi}, \citenamefont
  {Mahne}, \citenamefont {Manfredda}, \citenamefont {Medvedev}, \citenamefont
  {Nikolov}, \citenamefont {Pedersoli}, \citenamefont {Penco}, \citenamefont
  {Plekan}, \citenamefont {Prandolini}, \citenamefont {Prince}, \citenamefont
  {Raimondi}, \citenamefont {Rebernik}, \citenamefont {Riedel}, \citenamefont
  {Roussel}, \citenamefont {Sigalotti}, \citenamefont {Squibb}, \citenamefont
  {Stojanovic}, \citenamefont {Stranges}, \citenamefont {Svetina},
  \citenamefont {Tanikawa}, \citenamefont {Teubner}, \citenamefont {Tkachenko},
  \citenamefont {Toleikis}, \citenamefont {Zangrando}, \citenamefont {Ziaja},
  \citenamefont {Tavella},\ and\ \citenamefont
  {Giannessi}}]{finetti_pulse_2017}%
  \BibitemOpen
  \bibfield  {author} {\bibinfo {author} {\bibfnamefont {P.}~\bibnamefont
  {Finetti}}, \bibinfo {author} {\bibfnamefont {H.}~\bibnamefont {H\"oppner}},
  \bibinfo {author} {\bibfnamefont {E.}~\bibnamefont {Allaria}}, \bibinfo
  {author} {\bibfnamefont {C.}~\bibnamefont {Callegari}}, \bibinfo {author}
  {\bibfnamefont {F.}~\bibnamefont {Capotondi}}, \bibinfo {author}
  {\bibfnamefont {P.}~\bibnamefont {Cinquegrana}}, \bibinfo {author}
  {\bibfnamefont {M.}~\bibnamefont {Coreno}}, \bibinfo {author} {\bibfnamefont
  {R.}~\bibnamefont {Cucini}}, \bibinfo {author} {\bibfnamefont {M.~B.}\
  \bibnamefont {Danailov}}, \bibinfo {author} {\bibfnamefont {A.}~\bibnamefont
  {Demidovich}}, \bibinfo {author} {\bibfnamefont {G.}~\bibnamefont
  {De~Ninno}}, \bibinfo {author} {\bibfnamefont {M.}~\bibnamefont {Di~Fraia}},
  \bibinfo {author} {\bibfnamefont {R.}~\bibnamefont {Feifel}}, \bibinfo
  {author} {\bibfnamefont {E.}~\bibnamefont {Ferrari}}, \bibinfo {author}
  {\bibfnamefont {L.}~\bibnamefont {Fr\"ohlich}}, \bibinfo {author}
  {\bibfnamefont {D.}~\bibnamefont {Gauthier}}, \bibinfo {author}
  {\bibfnamefont {T.}~\bibnamefont {Golz}}, \bibinfo {author} {\bibfnamefont
  {C.}~\bibnamefont {Grazioli}}, \bibinfo {author} {\bibfnamefont
  {Y.}~\bibnamefont {Kai}}, \bibinfo {author} {\bibfnamefont {G.}~\bibnamefont
  {Kurdi}}, \bibinfo {author} {\bibfnamefont {N.}~\bibnamefont {Mahne}},
  \bibinfo {author} {\bibfnamefont {M.}~\bibnamefont {Manfredda}}, \bibinfo
  {author} {\bibfnamefont {N.}~\bibnamefont {Medvedev}}, \bibinfo {author}
  {\bibfnamefont {I.~P.}\ \bibnamefont {Nikolov}}, \bibinfo {author}
  {\bibfnamefont {E.}~\bibnamefont {Pedersoli}}, \bibinfo {author}
  {\bibfnamefont {G.}~\bibnamefont {Penco}}, \bibinfo {author} {\bibfnamefont
  {O.}~\bibnamefont {Plekan}}, \bibinfo {author} {\bibfnamefont {M.~J.}\
  \bibnamefont {Prandolini}}, \bibinfo {author} {\bibfnamefont {K.~C.}\
  \bibnamefont {Prince}}, \bibinfo {author} {\bibfnamefont {L.}~\bibnamefont
  {Raimondi}}, \bibinfo {author} {\bibfnamefont {P.}~\bibnamefont {Rebernik}},
  \bibinfo {author} {\bibfnamefont {R.}~\bibnamefont {Riedel}}, \bibinfo
  {author} {\bibfnamefont {E.}~\bibnamefont {Roussel}}, \bibinfo {author}
  {\bibfnamefont {P.}~\bibnamefont {Sigalotti}}, \bibinfo {author}
  {\bibfnamefont {R.}~\bibnamefont {Squibb}}, \bibinfo {author} {\bibfnamefont
  {N.}~\bibnamefont {Stojanovic}}, \bibinfo {author} {\bibfnamefont
  {S.}~\bibnamefont {Stranges}}, \bibinfo {author} {\bibfnamefont
  {C.}~\bibnamefont {Svetina}}, \bibinfo {author} {\bibfnamefont
  {T.}~\bibnamefont {Tanikawa}}, \bibinfo {author} {\bibfnamefont
  {U.}~\bibnamefont {Teubner}}, \bibinfo {author} {\bibfnamefont
  {V.}~\bibnamefont {Tkachenko}}, \bibinfo {author} {\bibfnamefont
  {S.}~\bibnamefont {Toleikis}}, \bibinfo {author} {\bibfnamefont
  {M.}~\bibnamefont {Zangrando}}, \bibinfo {author} {\bibfnamefont
  {B.}~\bibnamefont {Ziaja}}, \bibinfo {author} {\bibfnamefont
  {F.}~\bibnamefont {Tavella}}, \ and\ \bibinfo {author} {\bibfnamefont
  {L.}~\bibnamefont {Giannessi}},\ }\href {\doibase 10.1103/PhysRevX.7.021043}
  {\bibfield  {journal} {\bibinfo  {journal} {Physical Review X}\ }\textbf
  {\bibinfo {volume} {7}},\ \bibinfo {pages} {021043} (\bibinfo {year}
  {2017})}\BibitemShut {NoStop}%
\bibitem [{\citenamefont {St\"ohr}\ and\ \citenamefont
  {Siegmann}(2006)}]{stohr_magnetism:_2006}%
  \BibitemOpen
  \bibfield  {author} {\bibinfo {author} {\bibfnamefont {J.}~\bibnamefont
  {St\"ohr}}\ and\ \bibinfo {author} {\bibfnamefont {H.~C.}\ \bibnamefont
  {Siegmann}},\ }\href {//www.springer.com/gp/book/9783540302827} {\emph
  {\bibinfo {title} {Magnetism: {From} {Fundamentals} to {Nanoscale}
  {Dynamics}}}},\ Springer {Series} in {Solid}-{State} {Sciences}\ (\bibinfo
  {publisher} {Springer-Verlag},\ \bibinfo {address} {Berlin Heidelberg},\
  \bibinfo {year} {2006})\BibitemShut {NoStop}%
\bibitem [{\citenamefont {Signal}()}]{note3}%
  \BibitemOpen
  \bibfield  {author} {\bibinfo {author} {\bibnamefont {Signal}},\ }\href@noop
  {} {}\bibinfo {note} {As small as 0.7$\%$ can be measured for 500 shots which
  corresponds to an acquisition time of 50 seconds. Taking into account the CCD
  reading time and the set of 4 pictures and 2 magnetization, the total
  experimental time to get the XMCD signal is 20 minutes. This compare to a
  total experimental time of 12 minutes for single shot data.}\BibitemShut
  {Stop}%
\bibitem [{\citenamefont {Alebrand}\ \emph {et~al.}(2014)\citenamefont
  {Alebrand}, \citenamefont {Bierbrauer}, \citenamefont {Hehn}, \citenamefont
  {Gottwald}, \citenamefont {Schmitt}, \citenamefont {Steil}, \citenamefont
  {Fullerton}, \citenamefont {Mangin}, \citenamefont {Cinchetti},\ and\
  \citenamefont {Aeschlimann}}]{alebrand_subpicosecond_2014}%
  \BibitemOpen
  \bibfield  {author} {\bibinfo {author} {\bibfnamefont {S.}~\bibnamefont
  {Alebrand}}, \bibinfo {author} {\bibfnamefont {U.}~\bibnamefont
  {Bierbrauer}}, \bibinfo {author} {\bibfnamefont {M.}~\bibnamefont {Hehn}},
  \bibinfo {author} {\bibfnamefont {M.}~\bibnamefont {Gottwald}}, \bibinfo
  {author} {\bibfnamefont {O.}~\bibnamefont {Schmitt}}, \bibinfo {author}
  {\bibfnamefont {D.}~\bibnamefont {Steil}}, \bibinfo {author} {\bibfnamefont
  {E.~E.}\ \bibnamefont {Fullerton}}, \bibinfo {author} {\bibfnamefont
  {S.}~\bibnamefont {Mangin}}, \bibinfo {author} {\bibfnamefont
  {M.}~\bibnamefont {Cinchetti}}, \ and\ \bibinfo {author} {\bibfnamefont
  {M.}~\bibnamefont {Aeschlimann}},\ }\href {\doibase
  10.1103/PhysRevB.89.144404} {\bibfield  {journal} {\bibinfo  {journal}
  {Physical Review B}\ }\textbf {\bibinfo {volume} {89}},\ \bibinfo {pages}
  {144404} (\bibinfo {year} {2014})}\BibitemShut {NoStop}%
\bibitem [{\citenamefont {Medapalli}\ \emph {et~al.}(2012)\citenamefont
  {Medapalli}, \citenamefont {Razdolski}, \citenamefont {Savoini},
  \citenamefont {Khorsand}, \citenamefont {Kirilyuk}, \citenamefont {Kimel},
  \citenamefont {Rasing}, \citenamefont {Kalashnikova}, \citenamefont
  {Tsukamoto},\ and\ \citenamefont {Itoh}}]{medapalli_efficiency_2012}%
  \BibitemOpen
  \bibfield  {author} {\bibinfo {author} {\bibfnamefont {R.}~\bibnamefont
  {Medapalli}}, \bibinfo {author} {\bibfnamefont {I.}~\bibnamefont
  {Razdolski}}, \bibinfo {author} {\bibfnamefont {M.}~\bibnamefont {Savoini}},
  \bibinfo {author} {\bibfnamefont {A.~R.}\ \bibnamefont {Khorsand}}, \bibinfo
  {author} {\bibfnamefont {A.}~\bibnamefont {Kirilyuk}}, \bibinfo {author}
  {\bibfnamefont {A.~V.}\ \bibnamefont {Kimel}}, \bibinfo {author}
  {\bibfnamefont {T.}~\bibnamefont {Rasing}}, \bibinfo {author} {\bibfnamefont
  {A.~M.}\ \bibnamefont {Kalashnikova}}, \bibinfo {author} {\bibfnamefont
  {A.}~\bibnamefont {Tsukamoto}}, \ and\ \bibinfo {author} {\bibfnamefont
  {A.}~\bibnamefont {Itoh}},\ }\href {\doibase 10.1103/PhysRevB.86.054442}
  {\bibfield  {journal} {\bibinfo  {journal} {Physical Review B}\ }\textbf
  {\bibinfo {volume} {86}},\ \bibinfo {pages} {054442} (\bibinfo {year}
  {2012})}\BibitemShut {NoStop}%
\bibitem [{\citenamefont {Beaulieu}(2013)}]{beaulieu:tel-00953656}%
  \BibitemOpen
  \bibfield  {author} {\bibinfo {author} {\bibfnamefont {N.}~\bibnamefont
  {Beaulieu}},\ }\emph {\bibinfo {title} {{Electronic and magnetic properties
  under femtosecond laser excitation, from the Gd single crystal to the
  ferrimagnetic alloys}}},\ \href
  {https://tel.archives-ouvertes.fr/tel-00953656} {\bibinfo {type} {Theses}},\
  \bibinfo  {school} {{Universit{\'e} Paris Sud - Paris XI}} (\bibinfo {year}
  {2013})\BibitemShut {NoStop}%
\bibitem [{\citenamefont {Malinowski}\ \emph {et~al.}(2008)\citenamefont
  {Malinowski}, \citenamefont {Dalla~Longa}, \citenamefont {Rietjens},
  \citenamefont {Paluskar}, \citenamefont {Huijink}, \citenamefont {Swagten},\
  and\ \citenamefont {Koopmans}}]{malinowski_control_2008}%
  \BibitemOpen
  \bibfield  {author} {\bibinfo {author} {\bibfnamefont {G.}~\bibnamefont
  {Malinowski}}, \bibinfo {author} {\bibfnamefont {F.}~\bibnamefont
  {Dalla~Longa}}, \bibinfo {author} {\bibfnamefont {J.~H.~H.}\ \bibnamefont
  {Rietjens}}, \bibinfo {author} {\bibfnamefont {P.~V.}\ \bibnamefont
  {Paluskar}}, \bibinfo {author} {\bibfnamefont {R.}~\bibnamefont {Huijink}},
  \bibinfo {author} {\bibfnamefont {H.~J.~M.}\ \bibnamefont {Swagten}}, \ and\
  \bibinfo {author} {\bibfnamefont {B.}~\bibnamefont {Koopmans}},\ }\href
  {\doibase 10.1038/nphys1092} {\bibfield  {journal} {\bibinfo  {journal}
  {Nature Physics}\ }\textbf {\bibinfo {volume} {4}},\ \bibinfo {pages} {855}
  (\bibinfo {year} {2008})}\BibitemShut {NoStop}%
\bibitem [{\citenamefont {We}()}]{note2}%
  \BibitemOpen
  \bibfield  {author} {\bibinfo {author} {\bibnamefont {We}},\ }\href@noop {}
  {}\bibinfo {note} {Were able to measure a total time window of 1.2 ps, but
  because of border effect, IR leak, and shadows of higher diffraction order of
  the zone plate, the data are meaningful only from -0.2 to 0.6
  ps.}\BibitemShut {Stop}%
\bibitem [{\citenamefont {Fert\'e}\ \emph {et~al.}(2017)\citenamefont
  {Fert\'e}, \citenamefont {Bergeard}, \citenamefont {Le~Guyader},
  \citenamefont {Hehn}, \citenamefont {Malinowski}, \citenamefont {Terrier},
  \citenamefont {Otero}, \citenamefont {Holldack}, \citenamefont {Pontius},\
  and\ \citenamefont {Boeglin}}]{ferte_element-resolved_2017}%
  \BibitemOpen
  \bibfield  {author} {\bibinfo {author} {\bibfnamefont {T.}~\bibnamefont
  {Fert\'e}}, \bibinfo {author} {\bibfnamefont {N.}~\bibnamefont {Bergeard}},
  \bibinfo {author} {\bibfnamefont {L.}~\bibnamefont {Le~Guyader}}, \bibinfo
  {author} {\bibfnamefont {M.}~\bibnamefont {Hehn}}, \bibinfo {author}
  {\bibfnamefont {G.}~\bibnamefont {Malinowski}}, \bibinfo {author}
  {\bibfnamefont {E.}~\bibnamefont {Terrier}}, \bibinfo {author} {\bibfnamefont
  {E.}~\bibnamefont {Otero}}, \bibinfo {author} {\bibfnamefont
  {K.}~\bibnamefont {Holldack}}, \bibinfo {author} {\bibfnamefont
  {N.}~\bibnamefont {Pontius}}, \ and\ \bibinfo {author} {\bibfnamefont
  {C.}~\bibnamefont {Boeglin}},\ }\href {\doibase 10.1103/PhysRevB.96.134303}
  {\bibfield  {journal} {\bibinfo  {journal} {Physical Review B}\ }\textbf
  {\bibinfo {volume} {96}},\ \bibinfo {pages} {134303} (\bibinfo {year}
  {2017})}\BibitemShut {NoStop}%
\bibitem [{\citenamefont {Stamm}\ \emph {et~al.}(2007)\citenamefont {Stamm},
  \citenamefont {Kachel}, \citenamefont {Pontius}, \citenamefont {Mitzner},
  \citenamefont {Quast}, \citenamefont {Holldack}, \citenamefont {Khan},
  \citenamefont {Lupulescu}, \citenamefont {Aziz}, \citenamefont {Wietstruk},
  \citenamefont {D\"urr},\ and\ \citenamefont
  {Eberhardt}}]{stamm_femtosecond_2007}%
  \BibitemOpen
  \bibfield  {author} {\bibinfo {author} {\bibfnamefont {C.}~\bibnamefont
  {Stamm}}, \bibinfo {author} {\bibfnamefont {T.}~\bibnamefont {Kachel}},
  \bibinfo {author} {\bibfnamefont {N.}~\bibnamefont {Pontius}}, \bibinfo
  {author} {\bibfnamefont {R.}~\bibnamefont {Mitzner}}, \bibinfo {author}
  {\bibfnamefont {T.}~\bibnamefont {Quast}}, \bibinfo {author} {\bibfnamefont
  {K.}~\bibnamefont {Holldack}}, \bibinfo {author} {\bibfnamefont
  {S.}~\bibnamefont {Khan}}, \bibinfo {author} {\bibfnamefont {C.}~\bibnamefont
  {Lupulescu}}, \bibinfo {author} {\bibfnamefont {E.~F.}\ \bibnamefont {Aziz}},
  \bibinfo {author} {\bibfnamefont {M.}~\bibnamefont {Wietstruk}}, \bibinfo
  {author} {\bibfnamefont {H.~A.}\ \bibnamefont {D\"urr}}, \ and\ \bibinfo
  {author} {\bibfnamefont {W.}~\bibnamefont {Eberhardt}},\ }\href {\doibase
  10.1038/nmat1985} {\bibfield  {journal} {\bibinfo  {journal} {Nature
  Materials}\ }\textbf {\bibinfo {volume} {6}},\ \bibinfo {pages} {740}
  (\bibinfo {year} {2007})}\BibitemShut {NoStop}%
\bibitem [{\citenamefont {Boeglin}\ \emph {et~al.}(2010)\citenamefont
  {Boeglin}, \citenamefont {Beaurepaire}, \citenamefont {Halt\'e},
  \citenamefont {L\'opez-Flores}, \citenamefont {Stamm}, \citenamefont
  {Pontius}, \citenamefont {D\"urr},\ and\ \citenamefont
  {Bigot}}]{boeglin_distinguishing_2010}%
  \BibitemOpen
  \bibfield  {author} {\bibinfo {author} {\bibfnamefont {C.}~\bibnamefont
  {Boeglin}}, \bibinfo {author} {\bibfnamefont {E.}~\bibnamefont
  {Beaurepaire}}, \bibinfo {author} {\bibfnamefont {V.}~\bibnamefont
  {Halt\'e}}, \bibinfo {author} {\bibfnamefont {V.}~\bibnamefont
  {L\'opez-Flores}}, \bibinfo {author} {\bibfnamefont {C.}~\bibnamefont
  {Stamm}}, \bibinfo {author} {\bibfnamefont {N.}~\bibnamefont {Pontius}},
  \bibinfo {author} {\bibfnamefont {H.~A.}\ \bibnamefont {D\"urr}}, \ and\
  \bibinfo {author} {\bibfnamefont {J.-Y.}\ \bibnamefont {Bigot}},\ }\href
  {\doibase 10.1038/nature09070} {\bibfield  {journal} {\bibinfo  {journal}
  {Nature}\ }\textbf {\bibinfo {volume} {465}},\ \bibinfo {pages} {458}
  (\bibinfo {year} {2010})}\BibitemShut {NoStop}%
\bibitem [{\citenamefont {Koopmans}\ \emph {et~al.}(2010)\citenamefont
  {Koopmans}, \citenamefont {Malinowski}, \citenamefont {Dalla~Longa},
  \citenamefont {Steiauf}, \citenamefont {F\"ahnle}, \citenamefont {Roth},
  \citenamefont {Cinchetti},\ and\ \citenamefont
  {Aeschlimann}}]{koopmans_explaining_2010}%
  \BibitemOpen
  \bibfield  {author} {\bibinfo {author} {\bibfnamefont {B.}~\bibnamefont
  {Koopmans}}, \bibinfo {author} {\bibfnamefont {G.}~\bibnamefont
  {Malinowski}}, \bibinfo {author} {\bibfnamefont {F.}~\bibnamefont
  {Dalla~Longa}}, \bibinfo {author} {\bibfnamefont {D.}~\bibnamefont
  {Steiauf}}, \bibinfo {author} {\bibfnamefont {M.}~\bibnamefont {F\"ahnle}},
  \bibinfo {author} {\bibfnamefont {T.}~\bibnamefont {Roth}}, \bibinfo {author}
  {\bibfnamefont {M.}~\bibnamefont {Cinchetti}}, \ and\ \bibinfo {author}
  {\bibfnamefont {M.}~\bibnamefont {Aeschlimann}},\ }\href {\doibase
  10.1038/nmat2593} {\bibfield  {journal} {\bibinfo  {journal} {Nature
  Materials}\ }\textbf {\bibinfo {volume} {9}},\ \bibinfo {pages} {259}
  (\bibinfo {year} {2010})}\BibitemShut {NoStop}%
\bibitem [{\citenamefont {Battiato}\ \emph {et~al.}(2010)\citenamefont
  {Battiato}, \citenamefont {Carva},\ and\ \citenamefont
  {Oppeneer}}]{battiato_superdiffusive_2010}%
  \BibitemOpen
  \bibfield  {author} {\bibinfo {author} {\bibfnamefont {M.}~\bibnamefont
  {Battiato}}, \bibinfo {author} {\bibfnamefont {K.}~\bibnamefont {Carva}}, \
  and\ \bibinfo {author} {\bibfnamefont {P.~M.}\ \bibnamefont {Oppeneer}},\
  }\href {\doibase 10.1103/PhysRevLett.105.027203} {\bibfield  {journal}
  {\bibinfo  {journal} {Physical Review Letters}\ }\textbf {\bibinfo {volume}
  {105}},\ \bibinfo {pages} {027203} (\bibinfo {year} {2010})}\BibitemShut
  {NoStop}%
\bibitem [{\citenamefont {Pfau}\ \emph {et~al.}(2012)\citenamefont {Pfau},
  \citenamefont {Schaffert}, \citenamefont {M\"uller}, \citenamefont {Gutt},
  \citenamefont {Al-Shemmary}, \citenamefont {B\"uttner}, \citenamefont
  {Delaunay}, \citenamefont {D\"usterer}, \citenamefont {Flewett},
  \citenamefont {Fr\"omter}, \citenamefont {Geilhufe}, \citenamefont {Guehrs},
  \citenamefont {G\"unther}, \citenamefont {Hawaldar}, \citenamefont {Hille},
  \citenamefont {Jaouen}, \citenamefont {Kobs}, \citenamefont {Li},
  \citenamefont {Mohanty}, \citenamefont {Redlin}, \citenamefont {Schlotter},
  \citenamefont {Stickler}, \citenamefont {Treusch}, \citenamefont {Vodungbo},
  \citenamefont {Kl\"aui}, \citenamefont {Oepen}, \citenamefont {L\"uning},
  \citenamefont {Gr\"ubel},\ and\ \citenamefont
  {Eisebitt}}]{pfau_ultrafast_2012}%
  \BibitemOpen
  \bibfield  {author} {\bibinfo {author} {\bibfnamefont {B.}~\bibnamefont
  {Pfau}}, \bibinfo {author} {\bibfnamefont {S.}~\bibnamefont {Schaffert}},
  \bibinfo {author} {\bibfnamefont {L.}~\bibnamefont {M\"uller}}, \bibinfo
  {author} {\bibfnamefont {C.}~\bibnamefont {Gutt}}, \bibinfo {author}
  {\bibfnamefont {A.}~\bibnamefont {Al-Shemmary}}, \bibinfo {author}
  {\bibfnamefont {F.}~\bibnamefont {B\"uttner}}, \bibinfo {author}
  {\bibfnamefont {R.}~\bibnamefont {Delaunay}}, \bibinfo {author}
  {\bibfnamefont {S.}~\bibnamefont {D\"usterer}}, \bibinfo {author}
  {\bibfnamefont {S.}~\bibnamefont {Flewett}}, \bibinfo {author} {\bibfnamefont
  {R.}~\bibnamefont {Fr\"omter}}, \bibinfo {author} {\bibfnamefont
  {J.}~\bibnamefont {Geilhufe}}, \bibinfo {author} {\bibfnamefont
  {E.}~\bibnamefont {Guehrs}}, \bibinfo {author} {\bibfnamefont {C.~M.}\
  \bibnamefont {G\"unther}}, \bibinfo {author} {\bibfnamefont {R.}~\bibnamefont
  {Hawaldar}}, \bibinfo {author} {\bibfnamefont {M.}~\bibnamefont {Hille}},
  \bibinfo {author} {\bibfnamefont {N.}~\bibnamefont {Jaouen}}, \bibinfo
  {author} {\bibfnamefont {A.}~\bibnamefont {Kobs}}, \bibinfo {author}
  {\bibfnamefont {K.}~\bibnamefont {Li}}, \bibinfo {author} {\bibfnamefont
  {J.}~\bibnamefont {Mohanty}}, \bibinfo {author} {\bibfnamefont
  {H.}~\bibnamefont {Redlin}}, \bibinfo {author} {\bibfnamefont {W.~F.}\
  \bibnamefont {Schlotter}}, \bibinfo {author} {\bibfnamefont {D.}~\bibnamefont
  {Stickler}}, \bibinfo {author} {\bibfnamefont {R.}~\bibnamefont {Treusch}},
  \bibinfo {author} {\bibfnamefont {B.}~\bibnamefont {Vodungbo}}, \bibinfo
  {author} {\bibfnamefont {M.}~\bibnamefont {Kl\"aui}}, \bibinfo {author}
  {\bibfnamefont {H.~P.}\ \bibnamefont {Oepen}}, \bibinfo {author}
  {\bibfnamefont {J.}~\bibnamefont {L\"uning}}, \bibinfo {author}
  {\bibfnamefont {G.}~\bibnamefont {Gr\"ubel}}, \ and\ \bibinfo {author}
  {\bibfnamefont {S.}~\bibnamefont {Eisebitt}},\ }\href {\doibase
  10.1038/ncomms2108} {\bibfield  {journal} {\bibinfo  {journal} {Nature
  Communications}\ }\textbf {\bibinfo {volume} {3}},\ \bibinfo {pages} {1100}
  (\bibinfo {year} {2012})}\BibitemShut {NoStop}%
\bibitem [{\citenamefont {Vodungbo}\ \emph {et~al.}(2012)\citenamefont
  {Vodungbo}, \citenamefont {Gautier}, \citenamefont {Lambert}, \citenamefont
  {Sardinha}, \citenamefont {Lozano}, \citenamefont {Sebban}, \citenamefont
  {Ducousso}, \citenamefont {Boutu}, \citenamefont {Li}, \citenamefont {Tudu},
  \citenamefont {Tortarolo}, \citenamefont {Hawaldar}, \citenamefont
  {Delaunay}, \citenamefont {L\'opez-Flores}, \citenamefont {Arabski},
  \citenamefont {Boeglin}, \citenamefont {Merdji}, \citenamefont {Zeitoun},\
  and\ \citenamefont {L\"uning}}]{vodungbo_laser-induced_2012}%
  \BibitemOpen
  \bibfield  {author} {\bibinfo {author} {\bibfnamefont {B.}~\bibnamefont
  {Vodungbo}}, \bibinfo {author} {\bibfnamefont {J.}~\bibnamefont {Gautier}},
  \bibinfo {author} {\bibfnamefont {G.}~\bibnamefont {Lambert}}, \bibinfo
  {author} {\bibfnamefont {A.~B.}\ \bibnamefont {Sardinha}}, \bibinfo {author}
  {\bibfnamefont {M.}~\bibnamefont {Lozano}}, \bibinfo {author} {\bibfnamefont
  {S.}~\bibnamefont {Sebban}}, \bibinfo {author} {\bibfnamefont
  {M.}~\bibnamefont {Ducousso}}, \bibinfo {author} {\bibfnamefont
  {W.}~\bibnamefont {Boutu}}, \bibinfo {author} {\bibfnamefont
  {K.}~\bibnamefont {Li}}, \bibinfo {author} {\bibfnamefont {B.}~\bibnamefont
  {Tudu}}, \bibinfo {author} {\bibfnamefont {M.}~\bibnamefont {Tortarolo}},
  \bibinfo {author} {\bibfnamefont {R.}~\bibnamefont {Hawaldar}}, \bibinfo
  {author} {\bibfnamefont {R.}~\bibnamefont {Delaunay}}, \bibinfo {author}
  {\bibfnamefont {V.}~\bibnamefont {L\'opez-Flores}}, \bibinfo {author}
  {\bibfnamefont {J.}~\bibnamefont {Arabski}}, \bibinfo {author} {\bibfnamefont
  {C.}~\bibnamefont {Boeglin}}, \bibinfo {author} {\bibfnamefont
  {H.}~\bibnamefont {Merdji}}, \bibinfo {author} {\bibfnamefont
  {P.}~\bibnamefont {Zeitoun}}, \ and\ \bibinfo {author} {\bibfnamefont
  {J.}~\bibnamefont {L\"uning}},\ }\href {\doibase 10.1038/ncomms2007}
  {\bibfield  {journal} {\bibinfo  {journal} {Nature Communications}\ }\textbf
  {\bibinfo {volume} {3}},\ \bibinfo {pages} {999} (\bibinfo {year}
  {2012})}\BibitemShut {NoStop}%
\bibitem [{\citenamefont {La-O-Vorakiat}\ \emph {et~al.}(2012)\citenamefont
  {La-O-Vorakiat}, \citenamefont {Turgut}, \citenamefont {Teale}, \citenamefont
  {Kapteyn}, \citenamefont {Murnane}, \citenamefont {Mathias}, \citenamefont
  {Aeschlimann}, \citenamefont {Schneider}, \citenamefont {Shaw}, \citenamefont
  {Nembach},\ and\ \citenamefont {Silva}}]{la-o-vorakiat_ultrafast_2012}%
  \BibitemOpen
  \bibfield  {author} {\bibinfo {author} {\bibfnamefont {C.}~\bibnamefont
  {La-O-Vorakiat}}, \bibinfo {author} {\bibfnamefont {E.}~\bibnamefont
  {Turgut}}, \bibinfo {author} {\bibfnamefont {C.~A.}\ \bibnamefont {Teale}},
  \bibinfo {author} {\bibfnamefont {H.~C.}\ \bibnamefont {Kapteyn}}, \bibinfo
  {author} {\bibfnamefont {M.~M.}\ \bibnamefont {Murnane}}, \bibinfo {author}
  {\bibfnamefont {S.}~\bibnamefont {Mathias}}, \bibinfo {author} {\bibfnamefont
  {M.}~\bibnamefont {Aeschlimann}}, \bibinfo {author} {\bibfnamefont {C.~M.}\
  \bibnamefont {Schneider}}, \bibinfo {author} {\bibfnamefont {J.~M.}\
  \bibnamefont {Shaw}}, \bibinfo {author} {\bibfnamefont {H.~T.}\ \bibnamefont
  {Nembach}}, \ and\ \bibinfo {author} {\bibfnamefont {T.~J.}\ \bibnamefont
  {Silva}},\ }\href {\doibase 10.1103/PhysRevX.2.011005} {\bibfield  {journal}
  {\bibinfo  {journal} {Physical Review X}\ }\textbf {\bibinfo {volume} {2}},\
  \bibinfo {pages} {011005} (\bibinfo {year} {2012})}\BibitemShut {NoStop}%
\bibitem [{\citenamefont {Rudolf}\ \emph {et~al.}(2012)\citenamefont {Rudolf},
  \citenamefont {La-O-Vorakiat}, \citenamefont {Battiato}, \citenamefont
  {Adam}, \citenamefont {Shaw}, \citenamefont {Turgut}, \citenamefont
  {Maldonado}, \citenamefont {Mathias}, \citenamefont {Grychtol}, \citenamefont
  {Nembach}, \citenamefont {Silva}, \citenamefont {Aeschlimann}, \citenamefont
  {Kapteyn}, \citenamefont {Murnane}, \citenamefont {Schneider},\ and\
  \citenamefont {Oppeneer}}]{rudolf_ultrafast_2012}%
  \BibitemOpen
  \bibfield  {author} {\bibinfo {author} {\bibfnamefont {D.}~\bibnamefont
  {Rudolf}}, \bibinfo {author} {\bibfnamefont {C.}~\bibnamefont
  {La-O-Vorakiat}}, \bibinfo {author} {\bibfnamefont {M.}~\bibnamefont
  {Battiato}}, \bibinfo {author} {\bibfnamefont {R.}~\bibnamefont {Adam}},
  \bibinfo {author} {\bibfnamefont {J.~M.}\ \bibnamefont {Shaw}}, \bibinfo
  {author} {\bibfnamefont {E.}~\bibnamefont {Turgut}}, \bibinfo {author}
  {\bibfnamefont {P.}~\bibnamefont {Maldonado}}, \bibinfo {author}
  {\bibfnamefont {S.}~\bibnamefont {Mathias}}, \bibinfo {author} {\bibfnamefont
  {P.}~\bibnamefont {Grychtol}}, \bibinfo {author} {\bibfnamefont {H.~T.}\
  \bibnamefont {Nembach}}, \bibinfo {author} {\bibfnamefont {T.~J.}\
  \bibnamefont {Silva}}, \bibinfo {author} {\bibfnamefont {M.}~\bibnamefont
  {Aeschlimann}}, \bibinfo {author} {\bibfnamefont {H.~C.}\ \bibnamefont
  {Kapteyn}}, \bibinfo {author} {\bibfnamefont {M.~M.}\ \bibnamefont
  {Murnane}}, \bibinfo {author} {\bibfnamefont {C.~M.}\ \bibnamefont
  {Schneider}}, \ and\ \bibinfo {author} {\bibfnamefont {P.~M.}\ \bibnamefont
  {Oppeneer}},\ }\href {\doibase 10.1038/ncomms2029} {\bibfield  {journal}
  {\bibinfo  {journal} {Nature Communications}\ }\textbf {\bibinfo {volume}
  {3}},\ \bibinfo {pages} {1037} (\bibinfo {year} {2012})}\BibitemShut
  {NoStop}%
\bibitem [{\citenamefont {Bergeard}\ \emph {et~al.}(2014)\citenamefont
  {Bergeard}, \citenamefont {L\'opez-Flores}, \citenamefont {Halt\'e},
  \citenamefont {Hehn}, \citenamefont {Stamm}, \citenamefont {Pontius},
  \citenamefont {Beaurepaire},\ and\ \citenamefont
  {Boeglin}}]{bergeard_ultrafast_2014}%
  \BibitemOpen
  \bibfield  {author} {\bibinfo {author} {\bibfnamefont {N.}~\bibnamefont
  {Bergeard}}, \bibinfo {author} {\bibfnamefont {V.}~\bibnamefont
  {L\'opez-Flores}}, \bibinfo {author} {\bibfnamefont {V.}~\bibnamefont
  {Halt\'e}}, \bibinfo {author} {\bibfnamefont {M.}~\bibnamefont {Hehn}},
  \bibinfo {author} {\bibfnamefont {C.}~\bibnamefont {Stamm}}, \bibinfo
  {author} {\bibfnamefont {N.}~\bibnamefont {Pontius}}, \bibinfo {author}
  {\bibfnamefont {E.}~\bibnamefont {Beaurepaire}}, \ and\ \bibinfo {author}
  {\bibfnamefont {C.}~\bibnamefont {Boeglin}},\ }\href {\doibase
  10.1038/ncomms4466} {\bibfield  {journal} {\bibinfo  {journal} {Nature
  Communications}\ }\textbf {\bibinfo {volume} {5}} (\bibinfo {year} {2014}),\
  10.1038/ncomms4466}\BibitemShut {NoStop}%
\bibitem [{\citenamefont {Wieczorek}\ \emph {et~al.}(2015)\citenamefont
  {Wieczorek}, \citenamefont {Eschenlohr}, \citenamefont {Weidtmann},
  \citenamefont {R\"osner}, \citenamefont {Bergeard}, \citenamefont
  {Tarasevitch}, \citenamefont {Wehling},\ and\ \citenamefont
  {Bovensiepen}}]{wieczorek_separation_2015}%
  \BibitemOpen
  \bibfield  {author} {\bibinfo {author} {\bibfnamefont {J.}~\bibnamefont
  {Wieczorek}}, \bibinfo {author} {\bibfnamefont {A.}~\bibnamefont
  {Eschenlohr}}, \bibinfo {author} {\bibfnamefont {B.}~\bibnamefont
  {Weidtmann}}, \bibinfo {author} {\bibfnamefont {M.}~\bibnamefont {R\"osner}},
  \bibinfo {author} {\bibfnamefont {N.}~\bibnamefont {Bergeard}}, \bibinfo
  {author} {\bibfnamefont {A.}~\bibnamefont {Tarasevitch}}, \bibinfo {author}
  {\bibfnamefont {T.~O.}\ \bibnamefont {Wehling}}, \ and\ \bibinfo {author}
  {\bibfnamefont {U.}~\bibnamefont {Bovensiepen}},\ }\href {\doibase
  10.1103/PhysRevB.92.174410} {\bibfield  {journal} {\bibinfo  {journal}
  {Physical Review B}\ }\textbf {\bibinfo {volume} {92}},\ \bibinfo {pages}
  {174410} (\bibinfo {year} {2015})}\BibitemShut {NoStop}%
\bibitem [{\citenamefont {Jal}\ \emph {et~al.}(2017)\citenamefont {Jal},
  \citenamefont {L\'opez-Flores}, \citenamefont {Pontius}, \citenamefont
  {Fert\'e}, \citenamefont {Bergeard}, \citenamefont {Boeglin}, \citenamefont
  {Vodungbo}, \citenamefont {L\"uning},\ and\ \citenamefont
  {Jaouen}}]{jal_structural_2017}%
  \BibitemOpen
  \bibfield  {author} {\bibinfo {author} {\bibfnamefont {E.}~\bibnamefont
  {Jal}}, \bibinfo {author} {\bibfnamefont {V.}~\bibnamefont {L\'opez-Flores}},
  \bibinfo {author} {\bibfnamefont {N.}~\bibnamefont {Pontius}}, \bibinfo
  {author} {\bibfnamefont {T.}~\bibnamefont {Fert\'e}}, \bibinfo {author}
  {\bibfnamefont {N.}~\bibnamefont {Bergeard}}, \bibinfo {author}
  {\bibfnamefont {C.}~\bibnamefont {Boeglin}}, \bibinfo {author} {\bibfnamefont
  {B.}~\bibnamefont {Vodungbo}}, \bibinfo {author} {\bibfnamefont
  {J.}~\bibnamefont {L\"uning}}, \ and\ \bibinfo {author} {\bibfnamefont
  {N.}~\bibnamefont {Jaouen}},\ }\href {\doibase 10.1103/PhysRevB.95.184422}
  {\bibfield  {journal} {\bibinfo  {journal} {Physical Review B}\ }\textbf
  {\bibinfo {volume} {95}},\ \bibinfo {pages} {184422} (\bibinfo {year}
  {2017})}\BibitemShut {NoStop}%
\bibitem [{\citenamefont {Zhang}\ \emph {et~al.}(2017)\citenamefont {Zhang},
  \citenamefont {He}, \citenamefont {Zhang}, \citenamefont {Cheng},
  \citenamefont {Teng},\ and\ \citenamefont {F\"ahnle}}]{zhang_unifying_2017}%
  \BibitemOpen
  \bibfield  {author} {\bibinfo {author} {\bibfnamefont {W.}~\bibnamefont
  {Zhang}}, \bibinfo {author} {\bibfnamefont {W.}~\bibnamefont {He}}, \bibinfo
  {author} {\bibfnamefont {X.-Q.}\ \bibnamefont {Zhang}}, \bibinfo {author}
  {\bibfnamefont {Z.-H.}\ \bibnamefont {Cheng}}, \bibinfo {author}
  {\bibfnamefont {J.}~\bibnamefont {Teng}}, \ and\ \bibinfo {author}
  {\bibfnamefont {M.}~\bibnamefont {F\"ahnle}},\ }\href {\doibase
  10.1103/PhysRevB.96.220415} {\bibfield  {journal} {\bibinfo  {journal}
  {Physical Review B}\ }\textbf {\bibinfo {volume} {96}},\ \bibinfo {pages}
  {220415} (\bibinfo {year} {2017})}\BibitemShut {NoStop}%
\bibitem [{\citenamefont {Tengdin}\ \emph {et~al.}(2018)\citenamefont
  {Tengdin}, \citenamefont {You}, \citenamefont {Chen}, \citenamefont {Shi},
  \citenamefont {Zusin}, \citenamefont {Zhang}, \citenamefont {Gentry},
  \citenamefont {Blonsky}, \citenamefont {Keller}, \citenamefont {Oppeneer},
  \citenamefont {Kapteyn}, \citenamefont {Tao},\ and\ \citenamefont
  {Murnane}}]{tengdin_critical_2018}%
  \BibitemOpen
  \bibfield  {author} {\bibinfo {author} {\bibfnamefont {P.}~\bibnamefont
  {Tengdin}}, \bibinfo {author} {\bibfnamefont {W.}~\bibnamefont {You}},
  \bibinfo {author} {\bibfnamefont {C.}~\bibnamefont {Chen}}, \bibinfo {author}
  {\bibfnamefont {X.}~\bibnamefont {Shi}}, \bibinfo {author} {\bibfnamefont
  {D.}~\bibnamefont {Zusin}}, \bibinfo {author} {\bibfnamefont
  {Y.}~\bibnamefont {Zhang}}, \bibinfo {author} {\bibfnamefont
  {C.}~\bibnamefont {Gentry}}, \bibinfo {author} {\bibfnamefont
  {A.}~\bibnamefont {Blonsky}}, \bibinfo {author} {\bibfnamefont
  {M.}~\bibnamefont {Keller}}, \bibinfo {author} {\bibfnamefont {P.~M.}\
  \bibnamefont {Oppeneer}}, \bibinfo {author} {\bibfnamefont {H.~C.}\
  \bibnamefont {Kapteyn}}, \bibinfo {author} {\bibfnamefont {Z.}~\bibnamefont
  {Tao}}, \ and\ \bibinfo {author} {\bibfnamefont {M.~M.}\ \bibnamefont
  {Murnane}},\ }\href {\doibase 10.1126/sciadv.aap9744} {\bibfield  {journal}
  {\bibinfo  {journal} {Science Advances}\ }\textbf {\bibinfo {volume} {4}},\
  \bibinfo {pages} {eaap9744} (\bibinfo {year} {2018})}\BibitemShut {NoStop}%
\bibitem [{\citenamefont {You}\ \emph {et~al.}(2018)\citenamefont {You},
  \citenamefont {Tengdin}, \citenamefont {Chen}, \citenamefont {Shi},
  \citenamefont {Zusin}, \citenamefont {Zhang}, \citenamefont {Gentry},
  \citenamefont {Blonsky}, \citenamefont {Keller}, \citenamefont {Oppeneer},
  \citenamefont {Kapteyn}, \citenamefont {Tao},\ and\ \citenamefont
  {Murnane}}]{you_revealing_2018}%
  \BibitemOpen
  \bibfield  {author} {\bibinfo {author} {\bibfnamefont {W.}~\bibnamefont
  {You}}, \bibinfo {author} {\bibfnamefont {P.}~\bibnamefont {Tengdin}},
  \bibinfo {author} {\bibfnamefont {C.}~\bibnamefont {Chen}}, \bibinfo {author}
  {\bibfnamefont {X.}~\bibnamefont {Shi}}, \bibinfo {author} {\bibfnamefont
  {D.}~\bibnamefont {Zusin}}, \bibinfo {author} {\bibfnamefont
  {Y.}~\bibnamefont {Zhang}}, \bibinfo {author} {\bibfnamefont
  {C.}~\bibnamefont {Gentry}}, \bibinfo {author} {\bibfnamefont
  {A.}~\bibnamefont {Blonsky}}, \bibinfo {author} {\bibfnamefont
  {M.}~\bibnamefont {Keller}}, \bibinfo {author} {\bibfnamefont
  {P.}~\bibnamefont {Oppeneer}}, \bibinfo {author} {\bibfnamefont
  {H.}~\bibnamefont {Kapteyn}}, \bibinfo {author} {\bibfnamefont
  {Z.}~\bibnamefont {Tao}}, \ and\ \bibinfo {author} {\bibfnamefont
  {M.}~\bibnamefont {Murnane}},\ }\href {\doibase
  10.1103/PhysRevLett.121.077204} {\bibfield  {journal} {\bibinfo  {journal}
  {Physical Review Letters}\ }\textbf {\bibinfo {volume} {121}},\ \bibinfo
  {pages} {077204} (\bibinfo {year} {2018})}\BibitemShut {NoStop}%
\bibitem [{\citenamefont {Hansen}\ \emph {et~al.}(1991)\citenamefont {Hansen},
  \citenamefont {Klahn}, \citenamefont {Clausen}, \citenamefont {Much},\ and\
  \citenamefont {Witter}}]{hansen_magnetic_1991}%
  \BibitemOpen
  \bibfield  {author} {\bibinfo {author} {\bibfnamefont {P.}~\bibnamefont
  {Hansen}}, \bibinfo {author} {\bibfnamefont {S.}~\bibnamefont {Klahn}},
  \bibinfo {author} {\bibfnamefont {C.}~\bibnamefont {Clausen}}, \bibinfo
  {author} {\bibfnamefont {G.}~\bibnamefont {Much}}, \ and\ \bibinfo {author}
  {\bibfnamefont {K.}~\bibnamefont {Witter}},\ }\href {\doibase
  10.1063/1.348561} {\bibfield  {journal} {\bibinfo  {journal} {Journal of
  Applied Physics}\ }\textbf {\bibinfo {volume} {69}},\ \bibinfo {pages} {3194}
  (\bibinfo {year} {1991})}\BibitemShut {NoStop}%
\bibitem [{\citenamefont {Hohlfeld}\ \emph {et~al.}(1997)\citenamefont
  {Hohlfeld}, \citenamefont {Matthias}, \citenamefont {Knorren},\ and\
  \citenamefont {Bennemann}}]{Hohlfeld_nonequilibrium_1997}%
  \BibitemOpen
  \bibfield  {author} {\bibinfo {author} {\bibfnamefont {J.}~\bibnamefont
  {Hohlfeld}}, \bibinfo {author} {\bibfnamefont {E.}~\bibnamefont {Matthias}},
  \bibinfo {author} {\bibfnamefont {R.}~\bibnamefont {Knorren}}, \ and\
  \bibinfo {author} {\bibfnamefont {K.~H.}\ \bibnamefont {Bennemann}},\ }\href
  {\doibase 10.1103/PhysRevLett.78.4861} {\bibfield  {journal} {\bibinfo
  {journal} {Phys. Rev. Lett.}\ }\textbf {\bibinfo {volume} {78}},\ \bibinfo
  {pages} {4861} (\bibinfo {year} {1997})}\BibitemShut {NoStop}%
\bibitem [{\citenamefont {Roth}\ \emph {et~al.}(2012)\citenamefont {Roth},
  \citenamefont {Schellekens}, \citenamefont {Alebrand}, \citenamefont
  {Schmitt}, \citenamefont {Steil}, \citenamefont {Koopmans}, \citenamefont
  {Cinchetti},\ and\ \citenamefont {Aeschlimann}}]{Roth_temperature_2012}%
  \BibitemOpen
  \bibfield  {author} {\bibinfo {author} {\bibfnamefont {T.}~\bibnamefont
  {Roth}}, \bibinfo {author} {\bibfnamefont {A.~J.}\ \bibnamefont
  {Schellekens}}, \bibinfo {author} {\bibfnamefont {S.}~\bibnamefont
  {Alebrand}}, \bibinfo {author} {\bibfnamefont {O.}~\bibnamefont {Schmitt}},
  \bibinfo {author} {\bibfnamefont {D.}~\bibnamefont {Steil}}, \bibinfo
  {author} {\bibfnamefont {B.}~\bibnamefont {Koopmans}}, \bibinfo {author}
  {\bibfnamefont {M.}~\bibnamefont {Cinchetti}}, \ and\ \bibinfo {author}
  {\bibfnamefont {M.}~\bibnamefont {Aeschlimann}},\ }\href {\doibase
  10.1103/PhysRevX.2.021006} {\bibfield  {journal} {\bibinfo  {journal} {Phys.
  Rev. X}\ }\textbf {\bibinfo {volume} {2}},\ \bibinfo {pages} {021006}
  (\bibinfo {year} {2012})}\BibitemShut {NoStop}%
\bibitem [{\citenamefont {Atxitia}\ \emph {et~al.}(2010)\citenamefont
  {Atxitia}, \citenamefont {Chubykalo-Fesenko}, \citenamefont {Walowski},
  \citenamefont {Mann},\ and\ \citenamefont
  {M\"unzenberg}}]{Atxitia_evidence_2010}%
  \BibitemOpen
  \bibfield  {author} {\bibinfo {author} {\bibfnamefont {U.}~\bibnamefont
  {Atxitia}}, \bibinfo {author} {\bibfnamefont {O.}~\bibnamefont
  {Chubykalo-Fesenko}}, \bibinfo {author} {\bibfnamefont {J.}~\bibnamefont
  {Walowski}}, \bibinfo {author} {\bibfnamefont {A.}~\bibnamefont {Mann}}, \
  and\ \bibinfo {author} {\bibfnamefont {M.}~\bibnamefont {M\"unzenberg}},\
  }\href {\doibase 10.1103/PhysRevB.81.174401} {\bibfield  {journal} {\bibinfo
  {journal} {Phys. Rev. B}\ }\textbf {\bibinfo {volume} {81}},\ \bibinfo
  {pages} {174401} (\bibinfo {year} {2010})}\BibitemShut {NoStop}%
\bibitem [{\citenamefont {Suarez}\ \emph {et~al.}(2015)\citenamefont {Suarez},
  \citenamefont {Nieves}, \citenamefont {Laroze}, \citenamefont {Altbir},\ and\
  \citenamefont {Chubykalo-Fesenko}}]{suarez_ultrafast_2015}%
  \BibitemOpen
  \bibfield  {author} {\bibinfo {author} {\bibfnamefont {O.~J.}\ \bibnamefont
  {Suarez}}, \bibinfo {author} {\bibfnamefont {P.}~\bibnamefont {Nieves}},
  \bibinfo {author} {\bibfnamefont {D.}~\bibnamefont {Laroze}}, \bibinfo
  {author} {\bibfnamefont {D.}~\bibnamefont {Altbir}}, \ and\ \bibinfo {author}
  {\bibfnamefont {O.}~\bibnamefont {Chubykalo-Fesenko}},\ }\href {\doibase
  10.1103/PhysRevB.92.144425} {\bibfield  {journal} {\bibinfo  {journal}
  {Physical Review B}\ }\textbf {\bibinfo {volume} {92}},\ \bibinfo {pages}
  {144425} (\bibinfo {year} {2015})}\BibitemShut {NoStop}%
\bibitem [{\citenamefont {Atxitia}\ \emph {et~al.}(2014)\citenamefont
  {Atxitia}, \citenamefont {Barker}, \citenamefont {Chantrell},\ and\
  \citenamefont {Chubykalo-Fesenko}}]{Atxitia_controlling_2014}%
  \BibitemOpen
  \bibfield  {author} {\bibinfo {author} {\bibfnamefont {U.}~\bibnamefont
  {Atxitia}}, \bibinfo {author} {\bibfnamefont {J.}~\bibnamefont {Barker}},
  \bibinfo {author} {\bibfnamefont {R.~W.}\ \bibnamefont {Chantrell}}, \ and\
  \bibinfo {author} {\bibfnamefont {O.}~\bibnamefont {Chubykalo-Fesenko}},\
  }\href {\doibase 10.1103/PhysRevB.89.224421} {\bibfield  {journal} {\bibinfo
  {journal} {Phys. Rev. B}\ }\textbf {\bibinfo {volume} {89}},\ \bibinfo
  {pages} {224421} (\bibinfo {year} {2014})}\BibitemShut {NoStop}%
\bibitem [{\citenamefont {Moisan}\ \emph {et~al.}(2014)\citenamefont {Moisan},
  \citenamefont {Malinowski}, \citenamefont {Mauchain}, \citenamefont {Hehn},
  \citenamefont {Vodungbo}, \citenamefont {L\"uning}, \citenamefont {Mangin},
  \citenamefont {Fullerton},\ and\ \citenamefont
  {Thiaville}}]{moisan_investigating_2014}%
  \BibitemOpen
  \bibfield  {author} {\bibinfo {author} {\bibfnamefont {N.}~\bibnamefont
  {Moisan}}, \bibinfo {author} {\bibfnamefont {G.}~\bibnamefont {Malinowski}},
  \bibinfo {author} {\bibfnamefont {J.}~\bibnamefont {Mauchain}}, \bibinfo
  {author} {\bibfnamefont {M.}~\bibnamefont {Hehn}}, \bibinfo {author}
  {\bibfnamefont {B.}~\bibnamefont {Vodungbo}}, \bibinfo {author}
  {\bibfnamefont {J.}~\bibnamefont {L\"uning}}, \bibinfo {author}
  {\bibfnamefont {S.}~\bibnamefont {Mangin}}, \bibinfo {author} {\bibfnamefont
  {E.~E.}\ \bibnamefont {Fullerton}}, \ and\ \bibinfo {author} {\bibfnamefont
  {A.}~\bibnamefont {Thiaville}},\ }\href {\doibase 10.1038/srep04658}
  {\bibfield  {journal} {\bibinfo  {journal} {Scientific Reports}\ }\textbf
  {\bibinfo {volume} {4}} (\bibinfo {year} {2014}),\
  10.1038/srep04658}\BibitemShut {NoStop}%
\bibitem [{\citenamefont {Mendil}\ \emph {et~al.}(2014)\citenamefont {Mendil},
  \citenamefont {Nieves}, \citenamefont {Chubykalo-Fesenko}, \citenamefont
  {Walowski}, \citenamefont {Santos}, \citenamefont {Pisana},\ and\
  \citenamefont {M\"unzenberg}}]{mendil_resolving_2014}%
  \BibitemOpen
  \bibfield  {author} {\bibinfo {author} {\bibfnamefont {J.}~\bibnamefont
  {Mendil}}, \bibinfo {author} {\bibfnamefont {P.}~\bibnamefont {Nieves}},
  \bibinfo {author} {\bibfnamefont {O.}~\bibnamefont {Chubykalo-Fesenko}},
  \bibinfo {author} {\bibfnamefont {J.}~\bibnamefont {Walowski}}, \bibinfo
  {author} {\bibfnamefont {T.}~\bibnamefont {Santos}}, \bibinfo {author}
  {\bibfnamefont {S.}~\bibnamefont {Pisana}}, \ and\ \bibinfo {author}
  {\bibfnamefont {M.}~\bibnamefont {M\"unzenberg}},\ }\href {\doibase
  10.1038/srep03980} {\bibfield  {journal} {\bibinfo  {journal} {Scientific
  Reports}\ }\textbf {\bibinfo {volume} {4}} (\bibinfo {year} {2014}),\
  10.1038/srep03980}\BibitemShut {NoStop}%
\bibitem [{\citenamefont {Kuiper}\ \emph {et~al.}(2014)\citenamefont {Kuiper},
  \citenamefont {Roth}, \citenamefont {Schellekens}, \citenamefont {Schmitt},
  \citenamefont {Koopmans}, \citenamefont {Cinchetti},\ and\ \citenamefont
  {Aeschlimann}}]{kuiper_spin-orbit_2014}%
  \BibitemOpen
  \bibfield  {author} {\bibinfo {author} {\bibfnamefont {K.~C.}\ \bibnamefont
  {Kuiper}}, \bibinfo {author} {\bibfnamefont {T.}~\bibnamefont {Roth}},
  \bibinfo {author} {\bibfnamefont {A.~J.}\ \bibnamefont {Schellekens}},
  \bibinfo {author} {\bibfnamefont {O.}~\bibnamefont {Schmitt}}, \bibinfo
  {author} {\bibfnamefont {B.}~\bibnamefont {Koopmans}}, \bibinfo {author}
  {\bibfnamefont {M.}~\bibnamefont {Cinchetti}}, \ and\ \bibinfo {author}
  {\bibfnamefont {M.}~\bibnamefont {Aeschlimann}},\ }\href {\doibase
  10.1063/1.4902069} {\bibfield  {journal} {\bibinfo  {journal} {Applied
  Physics Letters}\ }\textbf {\bibinfo {volume} {105}},\ \bibinfo {pages}
  {202402} (\bibinfo {year} {2014})}\BibitemShut {NoStop}%
\end{thebibliography}%

\end{document}